\documentstyle[12pt,epsfig]{article}
\advance\textheight by \topskip
\begin{document}
\tolerance=100000
\thispagestyle{empty}
\setcounter{page}{0}
\def\lsim{\raisebox{-.1em}{$
\buildrel{\scriptscriptstyle <}\over{\scriptscriptstyle\sim}$}}
\def\gsim{\raisebox{-.1em}{$
\buildrel{\scriptscriptstyle >}\over{\scriptscriptstyle\sim}$}}
\def\preprint{{preprint}}
\begin{flushright}
{TUHEP-TH-02138}\\
{CERN-TH/2002-173}\\
{IPPP/02/47}\\
{DCPT/02/94}\\
{SHEP-03-03}\\
{CMS NOTE-2003/004}\\
\end{flushright}

\begin{center}
{\Large \bf
Trilepton$+$top signal from chargino-neutralino decays\\[0.15 cm]
of MSSM charged Higgs bosons at the LHC}\\[0.5 cm]
{\large Mike Bisset$^*$}\\[0.15 cm]
{\it Department of Physics, Tsinghua University,}\\
{\it Beijing, P.R. China 100084}
\\[0.5cm]
{\large Filip Moortgat$^*$}\\[0.15 cm]
{\it Department of Physics, University of Antwerpen,}\\
{\it B-2610 Antwerpen, Belgium}
\\[0.5cm]
{\large Stefano Moretti$^{*,\dagger}$}\\[0.15 cm]
{\it Department of Physics and Astronomy, University of Southampton,}\\
{\it Highfield, Southampton SO17 1BJ, UK}\\[0.25cm]
\end{center}
\vspace*{\fill}

\vskip -1.0cm

\begin{abstract}
{\vskip0.25cm\noindent\small 
We perform for the Large Hadron Collider (LHC) a detailed study of 
charged Higgs boson production via the top-bottom quark associated mode 
followed by decays into a chargino and a neutralino,
with masses and couplings as given by the general Minimal Supersymmetric
Standard Model (MSSM).
We focus our attention on the region of parameter space with
$m_{H^{\pm}} > m_t$ and intermediate values of $\tan\beta$,
where identification of $H^{\pm}$ via decays into
Standard Model (SM) particles  has proven to be ineffective.  
Modelling the CMS detector, we find that a signature consisting of 
three hard leptons accompanied by a hadronically reconstructed top quark
plus substantial missing transverse energy, which may result from
$H^{\pm} \rightarrow {\widetilde\chi}_{1,2}^{\pm}
{\widetilde\chi}_{1,2,3,4}^0$ decays, can be made viable over a large 
variety of 
initially overwhelming SM and MSSM backgrounds,
provided MSSM input parameters are favourable: notably, small
$| \mu |$ and light sleptons are important prerequisites.  
We quantify these statements by performing a fairly extensive scan of the
parameter space, including realistic hadron-level simulations, and
delineate some potential discovery regions.}
\end{abstract}
\vskip 0.25cm
\noindent
\hrule
\vskip 0.25cm
\noindent
\hskip0.00cm{$^*$E-mails: bisset@mail.tsinghua.edu.cn,
filip.moortgat@cern.ch, stefano.moretti@cern.ch.}\\
\noindent
\hskip0.00cm{$^\dagger$Formerly at: {\it Theory Division, CERN,}
{\it CH-1211 Geneva 23, Switzerland} {\rm and}
{\it Institute for Particle Physics Phenomenology,}
{\it University of Durham, Durham DH1 3LE, UK}.}
\vspace*{\fill}
\newpage

\noindent

\section{Introduction}

A pair of spin-less charged Higgs bosons, $H^{\pm}$
(with mass $m_{H^\pm}$), arises in any Two-Higgs Doublet Model (2HDM) 
alongside a trio of neutral Higgs bosons --- 
the $CP$-even `light' $h$ and `heavy' $H$ ({\it i.e.}, with $m_h < m_H$) 
scalars and the $CP$-odd pseudoscalar $\!A$ (with mass $m_A$).
Embedding a Type II 2HDM inside the attractive 
theoretical framework provided by Supersymmetry (SUSY) yields the 
MSSM (see \cite{guide}), wherein the particle
content is limited to the known SM states (fermions
and gauge bosons), their `sparticle' counterparts (sfermions and gauginos)
plus the five aforementioned Higgs bosons and their respective Higgsinos.
Among the new massive sparticles predicted in the MSSM are
the charginos and the neutralinos\footnote{We will refer to the
charginos and neutralinos collectively as `inos'.}, which are the mass 
eigenstate mixtures of the electroweak (EW) gauginos and the
Higgsinos.  Previous papers \cite{PAP1,PAP2} have
demonstrated that $H^{\pm}$ decays into a chargino and a neutralino can
probe regions of the MSSM
parameter space where charged Higgs boson decays into SM particles
and other Higgs bosons are swamped by backgrounds.
In particular,
$\tan\beta$ (the ratio of the vacuum expectation values
of the up-type and down-type Higgs doublets) values between $3$ and $10$
were found to be in part accessible via
$H^{\pm} \rightarrow
\widetilde{\chi}_1^{\pm} \widetilde{\chi}_{2,3}^0$
decay modes ({\it i.e.}, charged Higgs boson decays into the
lightest chargino and the second or third heaviest neutralino),
when the final state includes three leptons (meaning electrons and/or
muons)\footnote{The process is further identified by a hadronically
reconstructed top quark from the $tH^- X$ (or $\bar{t}H^+X$) production
process, and via substantial missing transverse momentum from the
lightest neutralinos, $\widetilde{\chi}_1^0$s, the stable 
Lightest Supersymmetric Particles (LSPs) which must eventually result from
decays of the inos.}.

Such $\tan\beta$ values fall in the so-called `intermediate' regime
wherein $H^{\pm}$ decays to SM objects (which may include neutral MSSM
Higgs bosons) are undetectable at the LHC irrespective of the values
chosen for other MSSM input parameters\footnote{Not coincidentally, in
roughly the same area coverage via the neutral Higgs sector is
questionable \cite{CMS,Filips}, particularly if the integrated luminosity
is limited (say, ${\sim}30\, \hbox{fb}^{-1}$).
In fact, the one neutral Higgs boson that may be detectable typically
mimics a SM Higgs boson (this is the so-called `decoupling scenario').}.
This zone of undetectability, in part due to the
$\sim(m_b^2\tan\beta^2 +m_t^2/\tan\beta^2)$ coupling 
of the main $pp\to tH^-X~+~{\mathrm{c.c.}}$ production 
mode\footnote{Charged Higgs bosons might also be produced in sparticle
decays \cite{cascades}.  In particular, decays of gluinos and/or squarks
could lead to copious numbers of $H^{\pm}$-containing events if these 
strongly-interacting sparticles are light enough to be abundantly 
produced.  
However, such events may also fail to contain a top quark or possess an 
excessive number of jets and so not satisfy our signal requirements.
Here we neglect such production processes.},
begins around $\tan\beta = 6$ or $7$ for
$m_{H^\pm}\sim m_t$ and spreads to encompass more and more $\tan\beta$
values (say, between $3$ and $20$) as $m_{H^\pm}$ grows larger. 
The rate suppression may be further exacerbated by the 
same $\tan\beta$ dependence in the $H^-\to b\bar t$ decays
if there are other competing decay channels --- naturally, if the
$H^-\to b\bar t$ branching ratio (BR) is $\simeq 1$, it will remain so
and there is no additional suppression because of the bottom-top decay
rate.  The alternative MSSM decay channel $H^-\to hW^-$, which 
also yields $b\bar b W^-$ intermediate states (since $h\to b\bar b$),
is only relevant within a minuscule $\tan\beta$ interval (roughly
$\tan\beta\approx 2-3$) for $m_{H^\pm}\, \lsim \, m_t$
--- this lies close to the LEP2's excluded region.
Then there is $H^-\to \tau\bar \nu_\tau$,
which is limited to larger $\tan\beta$ values\footnote{The 
$H^-\to s\bar c$ mode has a much reduced scope in
comparison, because of the large QCD background.},
at best offering coverage down to $\tan\beta \, \sim \, 10$
for $m_{H^\pm}\sim m_t$ and contracting to even higher $\tan\beta$ 
values as $m_{H^\pm}$ grows larger \cite{AssaCoa}.
(See references in \cite{PAP1,PAP2} for a list of
phenomenological analyses of these SM decay modes of a charged Higgs
boson.)

Considering such limitations, it is worthwhile pursuing further
the $H^{\pm} \rightarrow$ inos decay modes 
initially probed in \cite{PAP1,PAP2}\footnote{Hadron collider signals 
from neutral MSSM Higgs boson decays into inos were studied in 
\cite{Filips,neutstud}, while MSSM Higgs bosons BRs to inos, 
emphasising invisible decays to a pair of LSPs, were presented
in \cite{Early,GunHab}.  
},  
expanding upon the results found therein and placing the analysis
in a sounder phenomenological context.
The improvements found herein go in three general directions.
Firstly, the allowable parameter space is covered far more 
thoroughly, incorporating all possible chargino-neutralino 
decay modes into the analysis and including every conceivable path 
leading from a charged Higgs boson to a three leptons plus invisible 
energy final state.  Secondly, investigation
of the r\^ole of on- and off-shell sleptons (the SUSY partners of the
leptons) is considerably deepened: as noted in the previous studies, 
if there is a light slepton, the leptonic BRs of the
inos can be significantly enhanced (especially those of
$\widetilde{\chi}_2^0$ and/or $\widetilde{\chi}_3^0$).  
Thirdly, signals are herein studied within a full event
generator environment modelling the CMS detector
and also includes an improved background analysis that 
encompasses potential MSSM background processes
(\cite{PAP2} was a very preliminary account in both
these respects while \cite{PAP1} only considered SM backgrounds
and was carried out solely at the parton level).

The legacy of the CERN $e^+e^-$ collider is a model independent limit on 
$m_{H^\pm}$ from charged Higgs pair production of order $M_{W^\pm}$ ---
$78.6\, \hbox{GeV}$ is the current LEP2 bound \cite{HiggsLEP2}.
Further, the current lower Higgs boson mass bound of approximately 
$114\, \hbox{GeV}$  \cite{HiggsLEP2}
can be converted within the MSSM into a minimal value for 
$m_{H^{\pm}}$ of  ${\sim}130$--$140\, \hbox{GeV}$, for $\tan\!\beta \,
\simeq \,3$--$4$. 
This bound grows rapidly stronger as $\tan\!\beta$ is decreased while
tapering very gradually as $\tan\!\beta$ is increased  (staying in the
$110$--$125\, \hbox{GeV}$ interval for $\tan\beta~\gsim~6$).
For $m_{H^\pm}< m_t$, charged Higgs bosons could be discovered during
Run 2 of the FNAL Tevatron \cite{Run2}, which has 
already begun taking data at $\sqrt s_{p\bar p}= 2\, \hbox{TeV}$,
by exploiting their production in top and antitop quark decays 
($t\to b H^+ \; + \; \hbox{c.c.}$) followed by 
$H^- \to \tau^- \bar\nu_\tau\; + \; \hbox{c.c.}$ \cite{ioemono}.  
In contrast, if $m_{H^\pm} \, \gsim~m_t$ 
(our definition of a `heavy' charged Higgs boson), one will necessarily
have to wait until the advent of the LHC at CERN, with 
$\sqrt s_{pp}=14\, \hbox{TeV}$, and thus this study will concentrate 
upon charged Higgs boson masses well above that of the top (anti)quark.
This will also provide ample phase space to allow for decays into
sparticles with masses above current experimental bounds.

There are also other processes where charged Higgs bosons
(or $A$, to whose mass that of the $H^{\pm}$ is closely tied)
enter as virtual particles at the one-loop level.
These include neutral meson mixing 
($K^0 \bar{K}^0$, $D^0\bar{D}^0$ or $B^0 \bar{B}^0$) and
$Z \rightarrow b \bar{b}$ ($R_b$) \cite{loopconst},
$b \rightarrow s \gamma$ decays \cite{loopconst,bsgam},
$b \rightarrow c \tau \bar{\nu}_{\tau}$ decays \cite{bctau} 
and the anomalous muon magnetic dipole moment \cite{Gminus2}.
The $b \rightarrow s \gamma$ decays are
generally thought to be the most constraining \cite{loopconst}
($b \rightarrow c \tau \bar{\nu}_{\tau}$ becomes significant for 
very high values of $\tan\beta$).  In the MSSM, one-loop
diagrams with either a $H^{\pm}$--$t$ loop or a
$\widetilde{\chi}_1^{\pm}$--$\tilde{t}$ can give meaningful contributions
to $b \rightarrow s \gamma$ decay processes
(loops involving gluinos are not significant if gluinos are relatively
heavy), and these two contributions may come either with the same or
with opposite signs.  Thus $b \rightarrow s \gamma$ restrictions on
$m_{H^{\pm}}$ are linked to the masses of the lighter chargino and 
the stops.  The $b \rightarrow s \gamma$ decays and the other higher
order processes may well exclude some regions of the MSSM parameter space
that are still allowed by the more direct limits from Higgs boson
and sparticle searches at LEP2.  
However, definite bounds are quite difficult to delineate without restricting
oneself to some subset of the allowed parameter space of the general MSSM 
by specifying a mechanism for how SUSY is to be broken (and in general
linking together what in the general MSSM are independent input parameters).
Studies which have delineated excluded regions resulting from these
processes have invariably included additional assumptions about the
behaviour of the theory at higher energy scales --- such as in 
Minimal Supergravity (mSUGRA) for example, for which next-to-leading order
(NLO) calculations have recently been performed \cite{bsgam}.  There are
also significant uncertainties in translating the experimental results
into clear predictions about MSSM parameters \cite{FKN}.   
Concerning limits from recent $(g-2)_\mu$ measurements,
these are most restrictive \cite{Gminus2} when $\tan\beta$ is low
($\lsim~3$) -- a case which is not of particular interest for our process
-- and may be relaxed when smuons are light -- a case which is of
particular interest for our process. 

\section{MSSM Parameter Space}

Analysing the usefulness of  
$H^{\pm} \rightarrow$ chargino-neutralino decays within the 
general MSSM parameter space is a fairly involved undertaking since 
many independent input parameters associated with just about all the 
(s)particle sectors of the model can play crucial r\^oles.  
From the Higgs sector we of course have $\tan\beta$ along with 
one input Higgs boson mass, taken herein as $m_A$, to which the tree-level
masses of all the other Higgs bosons are pegged.  
These two inputs are largely sufficient for the SM decay modes,
assuming sparticle decay modes are closed.

Squark masses, particularly stop masses, can drive
significant radiative corrections to the tree-level Higgs boson masses,
especially to $m_h$. In contrast, higher order corrections to the 
tree-level relation $m_{H^{\pm}}^2 = m_A^2 + M_{W^\pm}^2$ are typically
quite small \cite{mhc-cor}.  Thus the signal rate is insensitive to the
choice of squark-sector inputs.  
Nevertheless, the coloured-sparticle sector affects the analysis in
peripheral --- but potentially crucial --- ways.
Firstly, the choice of the stop mass inputs can affect what regions of
the MSSM parameter space are excluded via Higgstrahlung or the
aforementioned $b \rightarrow s \gamma$ processes.  The former would
suggest choosing high stop inputs to help push $m_h$ up above the LEP2 
bounds, while the latter might prefer low stop inputs to cancel       
corrections due to a light chargino.  Be such arguments as they may, 
there is considerable uncertainty in the resulting limits on the 
general MSSM parameter space, and these issues will not be addressed
further.  The second consideration is the size of squark and gluino
backgrounds to our signature.  
Discussion of this will be postponed until the end of this section.

To specify the ino sector, the parameters $M_{2}$ and $\mu$, in addition to 
$\tan\beta$, are required.  $M_{1}$ is assumed to be determined
from $M_{2}$ via gaugino unification
({\it i.e.}, $M_{1} = \frac{5}{3}\tan^2\theta_W M_{2}$).  
This will determine the tree-level masses (to which the radiative corrections 
are quite modest) of the inos along with their couplings to the Higgs
bosons.  However, this is not enough, for the inos (except for 
$\widetilde{\chi}_1^0$) must also decay --
preferably into leptons for easy detection.  To calculate the leptonic
ino BRs, one must designate the properties of the slepton sector, since
light sleptons can greatly enhance said BRs \cite{PAP1,PAP2,BaerTata}.  
Inputs (assumed to be flavour-diagonal) from the slepton sector are the 
left and right soft slepton masses for each of the three generations
(selectrons, smuons, and staus) and the trilinear `$A$-terms' which come 
attached to Yukawa factors and thus only $A_{\tau}$ has a potential impact.
{\it A priori}, all six left and right mass inputs (and $A_{\tau}$) are
independent.  However, in most models currently advocated, one has
$m_{\tilde{e}_R} \simeq m_{\tilde{\mu}_R}$ and 
$m_{\tilde{e}_L} \simeq m_{\tilde{\mu}_L}$.  
We will assume such equalities to hold.  

\begin{figure}[!t]
\begin{center}
\epsfig{file=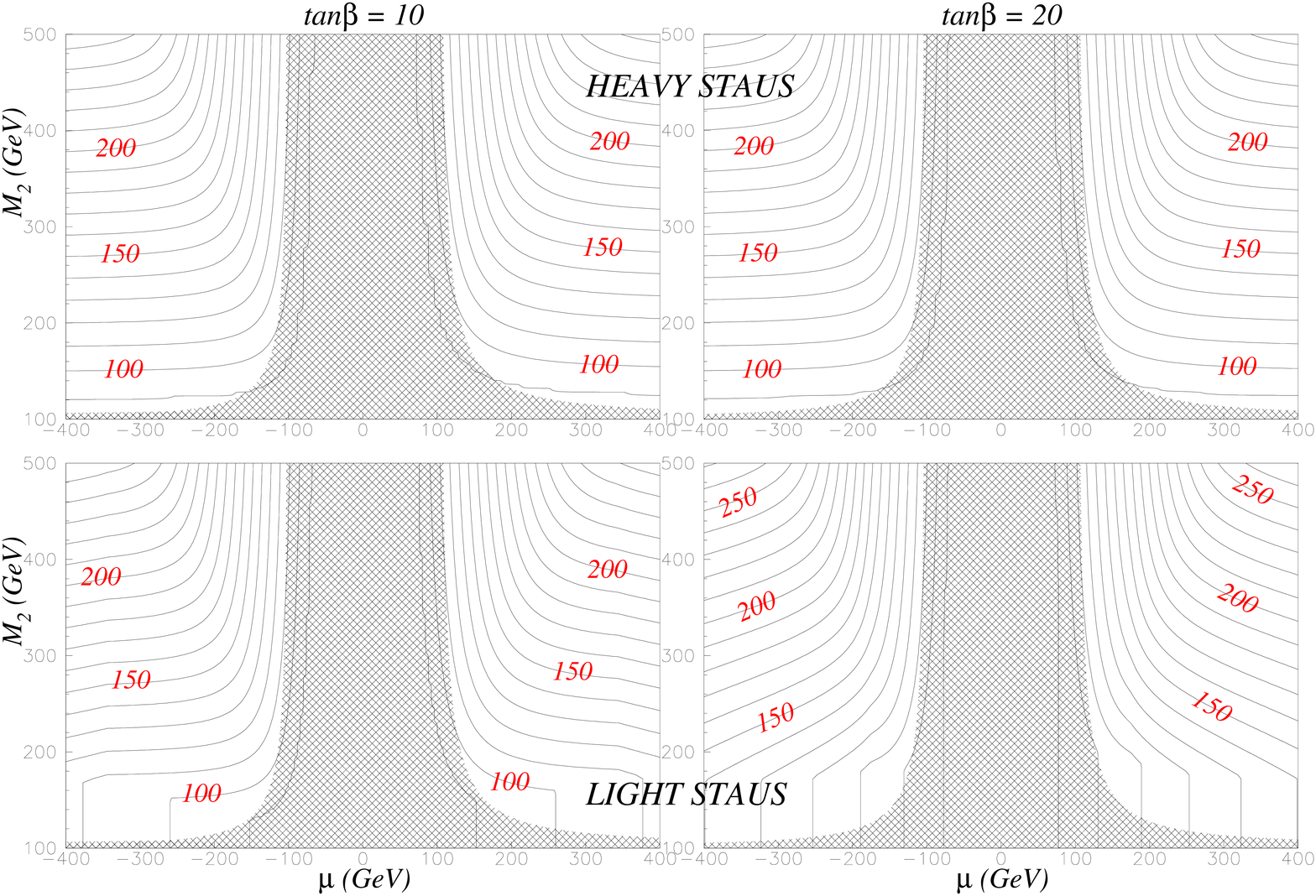,height=180mm, width=160mm}
\end{center}
\vskip -8.9cm
\caption{Minimum allowed soft slepton mass, given constraints as
described in the text.  In the upper (lower) two plots, 
soft stau mass inputs are set $100\, \hbox{GeV}$ above 
(degenerate with) those of the first two generations.
$A_{\ell} = 0$ in all cases. The shaded areas are excluded by LEP.}
\label{fig:slepmass1}
\end{figure}

To maximise leptonic ino BR enhancement, sleptons should be made as 
light as possible.  But direct searches at LEP2 \cite{W1LEP2} place
significant limits on slepton masses:
$m_{\tilde{e}_1} \ge 99.0\, \hbox{GeV}$,
$m_{\tilde{\mu}_1} \ge 91.0\, \hbox{GeV}$, 
$m_{\tilde{\tau}_1} \ge 85.0\, \hbox{GeV}$
(these assume that the slepton is not nearly-degenerate with the LSP)
and $m_{\tilde{\nu}} \ge 43.7\, \hbox{GeV}$
(from studies at the $Z$ pole).
Furthermore, the sneutrino masses are closely tied to the left soft mass
inputs, and, to avoid extra controversial
assumptions, we will restrict ourselves to regions of the MSSM parameter space 
where the LSP is the lightest neutralino rather than a sneutrino.
To optimise the ino leptonic BRs without running afoul of the
LEP2 limits, it is best to set
$m_{\tilde{\ell}_{\scriptscriptstyle R}} 
= m_{\tilde{\ell}_{\scriptscriptstyle L}}$.  
If all three generations have the same soft inputs
(with $A_\tau = 0$), then the  
slepton sector is effectively reduced to one optimal input value (which 
we identify with $m_{\tilde{\ell}_{\scriptscriptstyle R}}$).  
However, since ino decays to tau-leptons are generally not anywhere near 
as beneficial as are ino decays to electrons or muons,  it would be 
even better if the stau inputs were significantly above those of the 
first two generations.  This would enhance the inos' BRs into
electrons and muons.  In the general MSSM, we are of course free to choose
the inputs as such.  Doing so would also weaken restrictions from LEP2,
especially for high $\tan\beta$ values.  If we set the soft stau mass 
inputs $100\, \hbox{GeV}$ above those of the other sleptons (with $A_\tau$ 
still kept at zero), the lowest allowable slepton masses, presented
in the $M_2$ {\it vs.}\ $\mu$ plane for $\tan\beta = 10$ and $20$, are as
shown in the upper pair of plots in Fig.\ \ref{fig:slepmass1}, while if
all three generations have the same soft inputs we obtain the lower pair
of plots in Fig.\ \ref{fig:slepmass1}.

\begin{figure}[!t]
\begin{center}
\vskip -3.0cm
\epsfig{file=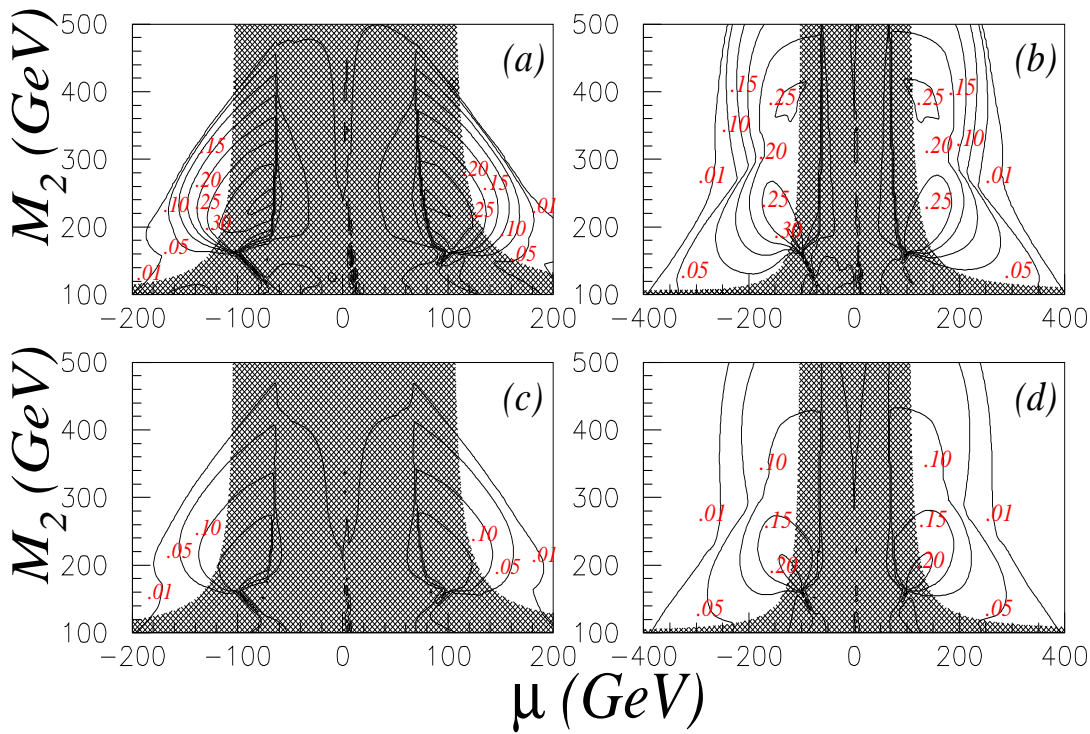,height=180mm, width=160mm}
\end{center}
\vskip -5.6cm
\caption{
BR$(H^{\pm} \rightarrow 3\ell N)$, where $\ell = e^{\pm}$ or
${\mu}^{\pm}$ and $N$ represents invisible final state particles,
with $(m_A, \, \tan\beta) =$
(a) $(300\, \hbox{GeV}, \, 10)$, 
(b) $(500\, \hbox{GeV}, \, 10)$,
(c) $(300\, \hbox{GeV}, \, 20)$,
(d) $(500\, \hbox{GeV}, \, 20)$.  
Slepton mass inputs are optimised as in 
the upper plots of Fig.\ \ref{fig:slepmass1}. 
The shaded areas are excluded by LEP;
$m_t = 175\, \hbox{GeV}$, $m_b = 4.25\, \hbox{GeV}$. 
}
\vskip -0.35cm
\label{fig:hto3ell-M2mu1}
\end{figure}

Incorporating such optimal slepton inputs and then 
scanning over the ino parameters $M_{2}$ and $\mu$, for a couple of values 
of $\tan\beta$ and $m_A$, yields Fig.\ \ref{fig:hto3ell-M2mu1} for 
BR$(H^{\pm} \rightarrow 3\ell N)$, where $\ell$ may be either $e^{\pm}$
or 
${\mu}^{\pm}$ and $N$ represents any number of undetectable final state
particles (either LSPs and/or neutrinos).
In these plots, and in plots to be shown hereafter, all possible
charged Higgs boson decay modes which can result in a final state with 
three charged leptons and no hadronic activity are included, except for
leptons coming from tau decays.  In this figure, including $\ell$s from
decaying taus would not noticeably affect the BRs since the staus which 
could greatly enhance tau production are pushed up in mass.

\begin{figure}[!t]
\begin{center}
\vskip-3.0cm
\epsfig{file=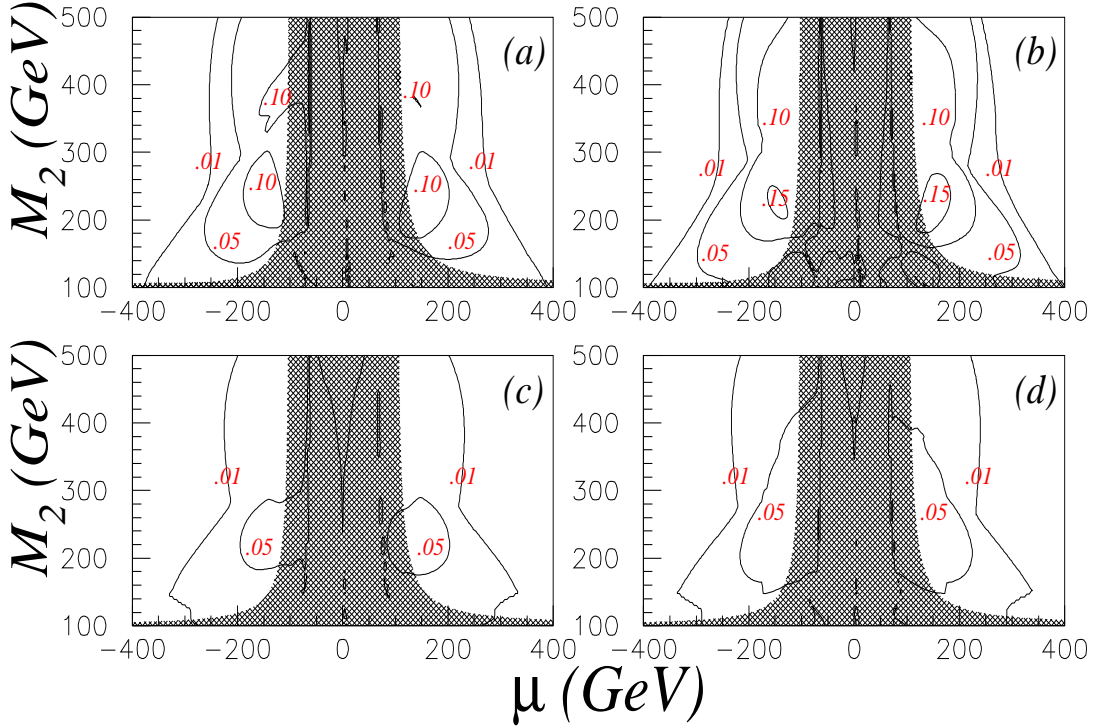,height=180mm, width=160mm} 
\end{center}
\vskip -5.5cm
\caption{
BR$(H^{\pm} \rightarrow 3\ell N)$, where $\ell = e^{\pm}$ or
${\mu}^{\pm}$ and $N$ represents invisible final state particles;  
$\ell$s resulting from tau decays 
(are not) are included for plots on the (left-) right-hand side;  
$m_A = 500\, \hbox{GeV}$ and
$\tan\beta = 10$ in plots (a) and (b), $20$ in plots (c) and (d).  
Slepton mass inputs for all three generations are optimised as in the
lower plots of Fig.\ \ref{fig:slepmass1}. 
The shaded areas are excluded by LEP;
$m_t = 175\, \hbox{GeV}$, $m_b = 4.25\, \hbox{GeV}$.
}
\vskip -0.3cm 
\label{fig:hto3ell-M2mu2}
\end{figure}

As expected, BRs are larger for the $m_A = 500\, \hbox{GeV}$ plots
on the right than for the $m_A = 300\, \hbox{GeV}$ plots on the left
since more $3\ell$-producing $H^{\pm} \rightarrow$ inos decay modes open 
up as $m_{H^{\pm}}$ increases.  BRs also decline as $\tan\beta$ is raised
from $10$ to $20$.  
If instead the three slepton generations have degenerate soft mass inputs, 
then Fig.\ \ref{fig:hto3ell-M2mu2} is obtained.  
Here $m_A$ is fixed at $500\, \hbox{GeV}$ and the left- and right-hand
plots depict, respectively, BRs without and with the inclusion of $\ell$s
from tau decays.  
Overall rates drop relative to those in Fig.\ \ref{fig:hto3ell-M2mu1}
since: (i) the slepton mass inputs must be set higher to evade LEP2
constraints; and (ii) charged Higgs boson decays leading to staus 
via inos -- which are now very significant -- often result in hadronic
final states rather than purely leptonic ones. 

Note from Figs.\ \ref{fig:hto3ell-M2mu1} \& \ref{fig:hto3ell-M2mu2}
that low values for $|{\mu}|$ are strongly favoured.  This can be
understood by inspecting the tree-level decay width formula 
for $H^{\pm} \rightarrow \widetilde{\chi}_i^{\pm} \widetilde{\chi}_j^0$
\cite{GunHab},
\begin{equation}\label{SUSYwidth}
\Gamma (H^{\pm} \rightarrow \widetilde\chi^{\pm}_i \widetilde\chi^0_j) =
\frac{g^2 \lambda^{1/2}[(F_L^2+F^2_R)
(m^2_{H^\pm} - m_{\widetilde\chi_i^\pm}^2 - m_{\widetilde\chi_j^0}^2) 
- 4\epsilon_j F_L F_R m_{\widetilde\chi_i^\pm}m_{\widetilde\chi_j^0}]}
{16 \pi m^3_{H^\pm}},
\end{equation}
\begin{eqnarray}
F_L &=& \cos\beta[N_{j4}V_{i1} + {\scriptstyle \sqrt{\frac{1}{2}}}
(N_{j2}+N_{j1}\tan\theta_W)V_{i2}], \nonumber \\  
F_R &=& \sin\beta[N_{j3}U_{i1} - {\scriptstyle \sqrt{\frac{1}{2}}}
(N_{j2}+N_{j1}\tan\theta_W)U_{i2}],
\label{eq:hdecay}
\end{eqnarray}
where $\epsilon_j$ is the sign for the neutralino mass eigenstate
(needed to generate a positive physical mass given the form of the 
neutralino mixing matrix),
$g$ is the 
$SU(2)_{\hbox{\smash{\lower 0.25ex \hbox{${\scriptstyle L}$}}}}$
coupling and 
$\lambda = (m^2_{H^\pm} - m_{\widetilde\chi_i^\pm}^2 -
m_{\widetilde\chi_j^0}^2)^2 - 4m_{\widetilde\chi_i^\pm}^2
m_{\widetilde\chi_j^0}^2$.
$V_{i1}$ \& $U_{i1}$ ($V_{i2}$ \& $U_{i2}$) give the gaugino
(Higgsino) component of chargino $\widetilde{\chi}_i^{\pm}$ 
while $N_{j1}$ \& $N_{j2}$ ($N_{j3}$ \& $N_{j4}$) give the gaugino
(Higgsino) components of neutralino $\widetilde{\chi}_j^0$.  
We immediately see from $F_L$ and $F_R$ that if the
chargino and the neutralino are both pure gauginos (the SUSY counterparts
of charged Higgs bosons decay into two gauge bosons --- for which there is 
no coupling at tree level) or both pure Higgsinos, then the tree-level 
decay width is zero.  Simple phase space considerations favouring
decays to lighter inos then disfavour situations in which 
$|\mu| \gg M_2$ (or $|\mu| \ll M_2$)
in which case light charginos {\sl and} light neutralinos are almost pure
gauginos (Higgsinos)
--- $|\mu| \sim M_2$ is preferred, ideally with both values as small as
possible to make the lighter inos as light as possible (to the extent that
LEP2 constraints permit).  Thus the optimal region for high 
$H^{\pm} \rightarrow$ inos BRs
is where inos are mixtures of gauginos and
Higgsinos just above the bends 
of the LEP2 parameter space bounds (shaded regions in Figs.\ 1--3) in the 
$M_2$ {\it vs.}\ $\mu$ plane\footnote{For higher values of $\tan\beta$,
$F_L\propto \cos\beta$ is small compared to $F_R\propto \sin\beta$.
So the $H^{\pm}$ to
$SU(2)_{\hbox{\smash{\lower 0.25ex \hbox{${\scriptstyle L}$}}}}$-wino
Higgsino decay SUSY-related to $H^{\pm} \rightarrow W^{\pm} h$
(where $h$ is now mostly 
from the
down-coupling Higgs doublet and so the corresponding Higgsino has a 
dominating $N_{j4}$ component entering into $F_L$) is also small.
But the actual inos may not have such compositions.  Furthermore,
the signature of $H^{\pm} \rightarrow$ inos is more distinctive than 
that of $H^{\pm} \rightarrow h W^{\pm}$ ---
even if the BRs for the two processes were similar, more events from
the former than from the latter would remain after sufficient cuts were
made to eliminate backgrounds.}.

In addition, one would like to optimise 
$H^{\pm} \rightarrow \widetilde{\chi}_i^{\pm} \widetilde{\chi}_j^0$
decays where $j \ne 1$ to obtain the vast majority of the 
decays generating three leptons.  Since $M_1 \simeq \frac{1}{2} M_2$,
$M_2\, \lsim \,|\mu|$ generates an LSP that is mostly a 
$U(1)_{\hbox{\smash{\lower 0.25ex \hbox{${\scriptstyle Y}$}}}}$
bino and a somewhat gaugino-dominated chargino --- which is bad for 
BR$(H^{\pm} \rightarrow \widetilde{\chi}_1^{\pm} \widetilde{\chi}_1^0$)
--- but also makes for a quite light LSP, which over-compensates for the 
sub-optimal coupling.  To increase the other $H^{\pm} \rightarrow$ light
inos BRs, the mass of the LSP may be raised by making $M_2$
somewhat larger than $|\mu|$.  Thus the final perscription for optimal
rates is for small $|\mu|$ values and slightly larger, but still small to
moderate values for $M_2$.

The charged Higgs boson BRs must now be tied to the 
production rate to obtain an expected number of signal events. 
Lowest order (LO) results from the parton-level process
$gb \rightarrow tH^{-}$ are strongly dependent on which
$b$-quark Parton Distribution Function (PDF) is chosen for
convolution and on the scale at which $\alpha_s$ is evaluated.
Moreover, the $b$-quark in the initial state originates with 
a gluon splitting into a $b\bar b$ pair inside the proton, so that the
above $2\to2$ process (when convoluted with initial state radiation
involving $g\to b\bar b$ in the backward evolution) can alternatively be
taken as the $2\to3$ hard scattering subprocess 
$gg \rightarrow \bar b t H^-$ interfaced to gluon PDFs.
The two descriptions have complementary strengths:
the former most aptly describes `inclusive' $tH^{-}X$ final states, 
as it re-sums to all orders large terms of the form
$\alpha_s \log(Q/m_b)$ 
(typically $Q \simeq m_t + M_{H^\pm}$, the choice we adopt here),
which are absorbed in the phenomenological PDF of the initial $b$-quark,
while the latter modelling is better at describing
`exclusive' observables, as it accounts for the correct kinematic
behaviour at large transverse momentum of the additional (or
spectator) $b$-quark in the final state. 
Yet contributions from the two processes cannot simply be summed.
In fact, the first term of the $b$-quark PDF is
given by the perturbative solution to the DGLAP equation
\vskip -0.2cm
\begin{equation}
b'(x,Q) = {\alpha_S \over \pi} \log \left({Q \over m_b}\right)
\int^1_x {dy \over y} P_{gb} \left({x \over y}\right) g(y,Q),
\label{seven}
\end{equation}
\vskip 0.1cm
\noindent
where $P_{gb} (z) = (z^2 + (1-z)^2)/2$ is the gluon-to-$b$ splitting
function, and the resulting contribution to $gb \rightarrow t H^-$ is
already accounted for by $gg \rightarrow \bar b t H^-$ in the
collinear limit.  Thus, when combining the $2\to2$ and $2\to3$ 
processes, the above contribution should be subtracted from the former to 
avoid double counting \cite{subtract}. 
An alternative approach \cite{Jaume} involves specifying a
threshold in the transverse momentum of the spectator $b$-quark,
$p_T^{b-{\mathrm{thr}}}$,
and then utilising $2\to2$ kinematics when $p_T^{b}<p_T^{b-{\mathrm{thr}}}$ 
and $2\to3$ kinematics for $p_T^{b}>p_T^{b-{\mathrm{thr}}}$.  This
is particularly well-suited to Monte Carlo (MC) event simulations 
since it does not involve making the aforementioned subtraction 
with its associated negative weights. 
Both techniques yield cross section values midway between the larger
predictions from $gb \rightarrow tH^{-}$ 
and the smaller ones from $gg \rightarrow \bar b t H^-$
(the latter being as much as a factor of 3--4 below the former). 
 
Both approaches are less sensitive to the choice of the $b$-quark PDF 
and the mass factorisation scale, $Q$, than if the two processes were
considered separately.
However, in each case, only some parts of the NLO corrections are 
accounted for, finally yielding a negative NLO contribution.  
Quite importantly, recent results \cite{ZhuPlehn} 
have proved that full NLO corrections to the $2\to2$ process 
({\it i.e.}, including both one-loop and radiative QCD corrections)
can yield an overall $K$-factor much larger than one\footnote{The size of 
the $K$-factor is sensitive to the choice of the mass factorisation scale,
$Q$, and the coupling renormalisation scale.  Equating these two scales
for convenience, $K$-factors near unity are found for low scale choices,
$Q \sim \frac{1}{8}(m_t + m_{H^{\pm}})$, while higher scale choices can
yield values on the order of $1.6$-$1.8$.  See \cite{ZhuPlehn} for further
details.  In all cases a $K$-factor much less than one is not found.},
overturning the negative corrections obtained via the above procedures.
Thus it is no longer justifiable to adopt normalisations
based on these techniques. 

The MSSM implementation \cite{MEs} of the HERWIG \cite{HERWIG} event
generator was used to simulate the $gb\to tH^-$ process and the 
various backgrounds.  As most backgrounds are only known to LO
accuracy, no additional $K$-factors were incorporated and
default LO PDFs and $\alpha_s$ values were employed.
This partly explains the improvement to be seen
herein relative to Ref.~\cite{PAP2}, where normalisation
was via the old subtraction procedure. 
Nonetheless, we still regard our results as conservative since
the dominant backgrounds (after cuts) are
$t\bar t$ production, which has a similar QCD $K$-factor
to that of the signal,  
and irreducible contributions from direct neutralino-chargino pair
production, which, being EW processes at tree level, have
smaller QCD corrections (of the order of 20\% or so \cite{PROSPINO}).
Yet one should also verify that the additional $b$-quark at high
transverse momentum produced by the $gg\to \bar b t H^-$ contribution, 
which is not present in the (infrared dominated) backward evolution of 
the $2\to2$ process' initial $b$-quark, does not render untrustworthy a
kinematical analysis done solely utilising the $2\to2$ process.  
We have confirmed this by also running HERWIG with $gg\to \bar{b}t H^-$ 
as the hard subprocess, adopting our usual selection cuts, and checking 
that in fact observable quantities (distributions and event rates) 
are not significantly affected by the presence of a spectator $b$-quark 
in the detector.    
All results shown will correspond to the outputs of the $2\to2$ process.

\begin{figure}[!t]
\begin{center}
\vskip -1.2cm
\epsfig{file=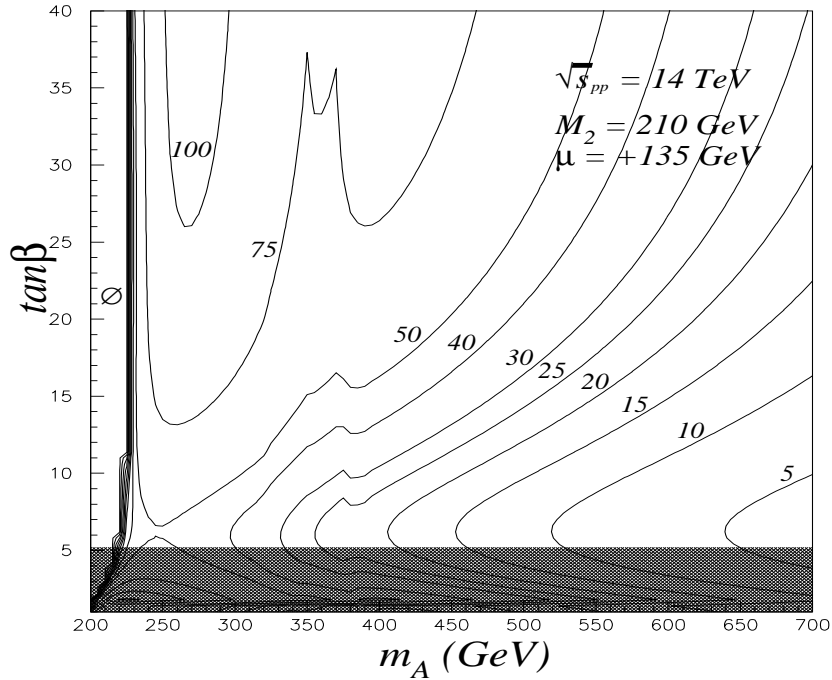,height=100mm, width=120mm}
\end{center}
\vskip -0.5cm
\caption{
$\sigma(pp \rightarrow tH^-X~+~{\rm{c.c.}})$ $\times$
BR$(H^{\pm} \rightarrow 3\ell N)$ (in fb), where $\ell = e^{\pm}$ or
${\mu}^{\pm}$ and $N$ represents invisible final state particles:
$M_{2}$ and $\mu$ are as noted, and
$\ell$s from tau decays are included.
Slepton mass inputs are optimised as in the upper plots of 
Fig.\ \ref{fig:slepmass1}.
The LEP2 $M_{\widetilde\chi_1^{\pm}}$ limit excludes
the shaded region, and the $\emptyset$ on the 
left signifies where the BR is virtually zero.}
\label{fig:hto3ell-tanBmA}
\vskip -0.3cm
\end{figure}

\begin{figure}[!t]
\begin{center}
\vskip -2.0cm
\epsfig{file=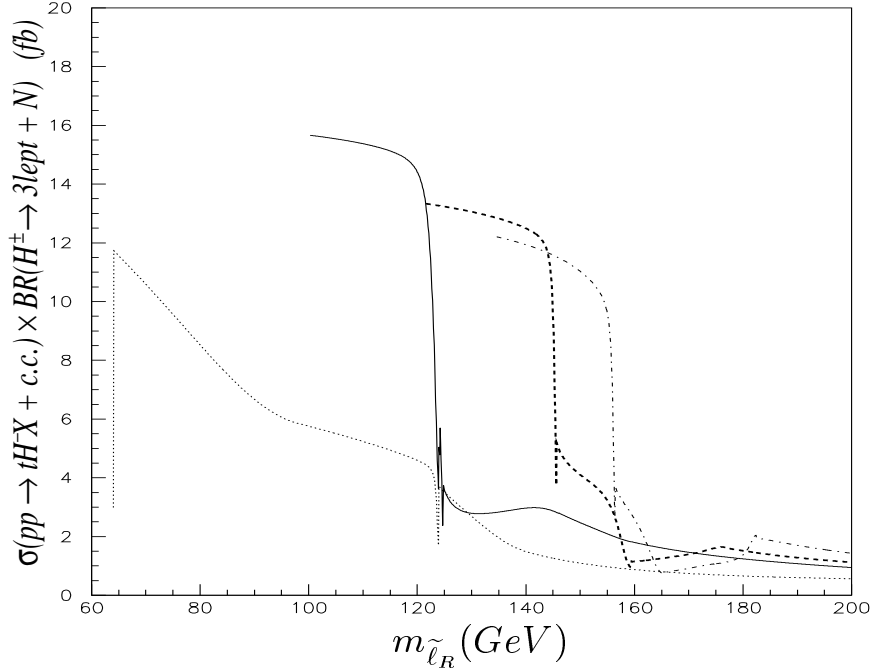,height=100mm, width=130mm}
\end{center}
\vskip -0.85cm
\caption{
$\sigma(pp \rightarrow tH^-X~+~{\rm{c.c.}})$ $\times$
BR$(H^{\pm} \rightarrow 3\ell N)$ (in fb), where 
$\ell = e^{\pm}$ or ${\mu}^{\pm}$
and $N$ represents invisible final state particles,
{\it vs.}\
$m_{\tilde{\ell}_{\scriptscriptstyle R}}$,
the soft slepton mass input for the first two
generations (soft stau mass inputs are pushed up by an additional 
$100\, \hbox{GeV}$, $A_{\tau}=0$). 
The set $(M_{2},\mu)$ is fixed at 
$(210\, \hbox{GeV}, +135\, \hbox{GeV})$
for Parameter Set A (solid curve), at
$(280\, \hbox{GeV}, +150\, \hbox{GeV})$  
for Parameter Set B (thick dashed curve) and at
$(300\, \hbox{GeV}, -150\, \hbox{GeV})$  
for Parameter Set C (dot-dashed curve).
The dotted curve replaces
$m_{\tilde{\ell}_{\scriptscriptstyle L}} 
= m_{\tilde{\ell}_{\scriptscriptstyle R}}$ in
Parameter Set A with
$m_{\tilde{\ell}_{\scriptscriptstyle L}} 
= m_{\tilde{\ell}_{\scriptscriptstyle R}} \, + \, 100\, \hbox{GeV}$.
The curves are terminated at the left where they would be  
LEP2 excluded, including the additional condition that
$m_{\tilde\nu} > m_{\widetilde\chi_1^0}$;
$\ell$s from tau decays are included.  
}
\label{fig:hto3ell-mslep}
\vskip -0.3cm
\end{figure}

Fig.\ \ref{fig:hto3ell-tanBmA} shows 
$\sigma(pp \rightarrow tH^-X~+~{\mathrm{c.c.}})$ with the
subsequent decay $H^{\pm} \rightarrow 3\ell N$, where 
$M_{2}$ and $\mu$ are fixed at the favourable values of 
$210\, \hbox{GeV}$ and $135\, \hbox{GeV}$, respectively
--- leading to the exclusion of $\tan\beta$ values below ${\sim}~5$   
by the $103\, \hbox{GeV}$ LEP2 lower bound on the chargino mass
\cite{W1LEP2}\footnote{For this choice of input parameters, the 
$m_{\widetilde\chi_1^\pm}$ bound is {\sl probably} more restrictive than
the one from Higgstrahlung, $e^+e^- \rightarrow h Z$ (and $hA$); however,
this will not be true for other choices of $M_2$ and $\mu$, such as those
considered in the next paragraph.  Note that the location of the 
Higgstrahlung bound is quite vague due to uncertainties in
the radiatively-corrected mass $m_h$ and errors in the measured value of
$m_t$.}. 
In the plot, the preference for high and low values of $\tan\beta$
so well-known for the raw $pp \rightarrow tH^-X$ cross section 
remains, though rates are nevertheless sufficient to seek a visible signal
even in the intermediate $\tan\beta$ region via our characteristic
signature.

It is instructive to next isolate the dependence of the signal rate
upon the masses of the sleptons\footnote{Though 
sleptons are light, direct $H^{\pm}$ BRs to slepton
pairs are at the sub-percent level.  Sleptons meaningfully 
influence charged Higgs boson leptonic BRs via the sleptons' 
involvement in subsequent ino decays.}.  
This is done in Fig.\ \ref{fig:hto3ell-mslep} for 
three choices of the other relevant MSSM parameters.  All combinations
fix $\tan\beta$ at $10$ and $m_A$ at $500\, \hbox{GeV}$.  
Parameter Set A
(solid curve in Fig.\ \ref{fig:hto3ell-mslep}) also sets 
$\mu = 135\, \hbox{GeV}$ and $M_{2} = 210\, \hbox{GeV}$ 
(as in Fig.\ \ref{fig:hto3ell-tanBmA}), 
while Parameter Set B 
(thick dashed curve in Fig.\ \ref{fig:hto3ell-mslep}) has 
$\mu = 150\, \hbox{GeV}$ and $M_{2} = 280\, \hbox{GeV}$ 
and Parameter Set C (dot-dashed curve in Fig.\ \ref{fig:hto3ell-mslep})
adopts $\mu = -150\, \hbox{GeV}$ and $M_{2} = 300\, \hbox{GeV}$ 
(these same parameter sets will also be used in the forthcoming 
experimental analysis).  
The horizontal axis in Fig.\ \ref{fig:hto3ell-mslep} is the 
{\sl soft} slepton mass input (as before, left and right soft masses are 
degenerate and $A$-terms are zero).  Bear in mind
that this is not the same as the physical masses of the various sleptons, 
which also have so-called $D$-term contributions.  
The curves are terminated on the left side at the point where LEP
experiments preclude the resulting light sleptons.  
Also shown by the dotted curve is the effect of removing
the equality $m_{\tilde{\ell}_{\scriptscriptstyle L}} 
= m_{\tilde{\ell}_{\scriptscriptstyle R}}$:  in this case
$m_{\tilde{\ell}_{\scriptscriptstyle L}} 
= m_{\tilde{\ell}_{\scriptscriptstyle R}} + 100\,~\hbox{GeV}$
while all the other MSSM parameters are the same as in Parameter Set A.  

Focusing the account upon the two curves relating to Parameter 
Set A (features of the other curves are seen to be qualitatively
similar), a sharp drop is seen around
$m_{\tilde{\ell}_{\scriptscriptstyle R}} \sim 123-125\, \hbox{GeV}$
where the second neutralino becomes degenerate with the charged 
selectrons and smuons and also where the lighter chargino 
becomes degenerate with the sneutrinos (of the first two
generations).  The drop is due to the closing of the 
two-body decay modes
$\widetilde{\chi}_2^0 \rightarrow \tilde{\ell}^{\pm} \ell^{\mp}$
and
$\widetilde{\chi}_1^+ \rightarrow \tilde{\nu}_\ell \ell^+$,
where $\tilde{\ell}^{\pm}$ and $\tilde{\nu}_\ell$ are 
on-mass-shell.  Although the two-body decay modes close at this point, 
the sleptons still make their presence felt in the associated three-body
decay modes via off-shell contributions.  
A modest rise in the rate occurs for the solid curve as
$m_{\tilde{\ell}_{\scriptscriptstyle R}}$ reaches ${\sim}146\, \hbox{GeV}$
where the second neutralino and the sneutrinos
(of the first two generations) become degenerate and the `spoiler' modes,
$\widetilde{\chi}_2^0 \rightarrow \tilde{\nu}_\ell \nu_\ell$ --- which
result in no charged leptons --- become inaccessible, consequently 
allowing BRs for the $\ell$-producing channels to rise.  
This feature is absent for the dotted curve, with
$m_{\tilde{\ell}_{\scriptscriptstyle L}} =
m_{\tilde{\ell}_{\scriptscriptstyle R}} + 100\, \hbox{GeV}$, since
two-body ino decay modes to the now too heavy sneutrinos are not open.  
Yet despite the absence of the spoiler modes, 
rates remain lower in this case because $\tilde{\ell}_L$s are also
heavy, simultaneously weakening the rate to charged leptons
(when both are accessible and neither is phase-space suppressed, 
ino decays to either $\tilde{\ell}_L$s or to $\tilde{\nu}_{\ell}$s 
may be larger, depending on the composition of the neutralino).  
Note also that the dip at 
$m_{\tilde{\ell}_{\scriptscriptstyle R}} \sim 123-125\, \hbox{GeV}$ 
is less pronounced since now only two-body decay modes to 
$\tilde{\ell}_R$s are turning off at this point rather than to both
left and right charged sleptons as in the solid curve.  
Lower values of $m_{\tilde{\ell}_{\scriptscriptstyle R}}$ are
now possible since our (perhaps unnecessarily restrictive) requirement 
that $m_{\tilde\nu} > m_{\widetilde\chi_1^0}$ is satisfied for
lower $m_{\tilde{\ell}_{\scriptscriptstyle R}}$ values.  However, even
going to such modest $m_{\tilde{\ell}_{\scriptscriptstyle R}}$ values does
not compensate for the enhancement obtained when   
$m_{\tilde{\ell}_{\scriptscriptstyle L}} =   
m_{\tilde{\ell}_{\scriptscriptstyle R}}$ and the two-body modes to
left charged sleptons are available.
Thus the peak magnitude is appreciably lower for the dotted curve
(confirming that $m_{\tilde{\ell}_{\scriptscriptstyle L}} =
m_{\tilde{\ell}_{\scriptscriptstyle R}}$ is the optimal
setup)\footnote{Setting
$m_{\tilde{\ell}_{\scriptscriptstyle R}}   
> m_{\tilde{\ell}_{\scriptscriptstyle L}}$ shifts the curve to the right
and slightly lowers the peak plateau.}.  

For Parameter Set A, with $m_{\tilde{\ell}_{\scriptscriptstyle R}}$
set to $110\, \hbox{GeV}$, the largest contributor to the signal events 
is in fact\footnote{Numbers include leptons from decaying taus,
but said inclusion only causes slight changes.} 
$H^{\pm} \rightarrow \widetilde{\chi}_1^{\pm}
\widetilde{\chi}_2^0$ (35.2\%), followed closely by 
$H^{\pm} \rightarrow \widetilde{\chi}_2^{\pm} 
\widetilde{\chi}_2^0$ (34.8\%) and then 
$H^{\pm} \rightarrow \widetilde{\chi}_1^{\pm}
\widetilde{\chi}_4^0$ (19.3\%), 
$H^{\pm} \rightarrow \widetilde{\chi}_2^{\pm}
\widetilde{\chi}_3^0$ (6.5\%),
and small contributions from  
$H^{\pm} \rightarrow \widetilde{\chi}_1^{\pm}
\widetilde{\chi}_3^0$ (3.0\%)
and
$H^{\pm} \rightarrow \widetilde{\chi}_2^{\pm}
\widetilde{\chi}_1^0$ (1.2\%)
(here, the three $\ell$s all come from the 
$\widetilde{\chi}_2^{\pm}$); $H^{\pm} \rightarrow \widetilde{\chi}_1^{\pm}
\widetilde{\chi}_{2,3}^0$ channels do not lead to most of the 
prospective signal events, contrary to what was assumed in Ref.~\cite{PAP1}.  
Nevertheless, rates seen in Fig.\ \ref{fig:hto3ell-mslep}
are still closely linked to $m_{\widetilde\chi_1^\pm}$ and 
$m_{\widetilde\chi_2^0}$ since $\widetilde{\chi}_1^{\pm}$ and/or
$\widetilde{\chi}_2^0$ are present in most (92.3\% for Set A) events, and,
even if one (or both) is not in the ino pair to which $H^{\pm}$
directly decays, the heavier inos into which $H^{\pm}$ does decay
in turn sometimes decay into these lighter inos (and charged leptons or
neutrinos) to generate the signal events.
With so many contributing channels, some of which involve
multiple sparticle to sparticle decay chains, simulation of the
signal with a robust event generator is imperative to ascertain the 
percentage of the events predicted utilising sparticle BR assignments 
that survive the cuts needed to sufficiently identify the signature and
eliminate the backgrounds.  

Returning now to the question of potential backgrounds from
coloured-sparticle production processes,
gluinos and squarks of the first two generations {\sl may} in principle 
produce multi-lepton events with top quarks; however, in practice, 
top quarks are quite often not present in such events.  Further, 
the limit on the squark (gluino) masses from Tevatron studies is now
at least ${\sim}260\, \hbox{GeV}$ (${\sim}190\, \hbox{GeV}$) \cite{PDBTev}, 
and will rise if Tevatron searches continue to be unsuccessful.  
In addition, if the gaugino unification assumption
also encompasses the gluino, then the gluino mass would be in the range
${\sim}700$--$1000\, \hbox{GeV}$ for the points being considered, and
(at least in mSUGRA-inspired scenarios) squarks are expected 
to have heavier or at least comparable masses \cite{colorvac}.  
Thus there is substantial rationale for limiting this analysis to 
heavy gluino and squarks (of the first two generations) masses.

Stops are different though.  Stringent experimental limits from LEP2 on  
stop masses only set a lower bound of 
${\sim}100\, \hbox{GeV}$ \cite{W1LEP2}, and  
stop pair production will generally lead to events containing top quarks.
Possible decay chains that could mimic our signal events include for
example
$\widetilde{t} \widetilde{t}^* \rightarrow
\widetilde{\chi}_1^0 (t \rightarrow b \ell \nu)
\; + \;
( \widetilde{\chi}_2^0 \rightarrow \widetilde{\chi}_1^0 \ell \ell )
(t \rightarrow \hbox{hadrons})$ and
$\widetilde{t} \widetilde{t}^* \rightarrow
(\widetilde{\chi}_1^+ \rightarrow \widetilde{\chi}_1^0 \ell \nu)  b
\; + \;
( \widetilde{\chi}_2^0 \rightarrow \widetilde{\chi}_1^0 \ell \ell )
(t \rightarrow \hbox{hadrons})$.
Sbottom pair production can also yield such a final state, as for 
example via 
$\widetilde{b} \widetilde{b}^* \rightarrow
(\widetilde{\chi}_1^+ \rightarrow \widetilde{\chi}_1^0 \ell \nu)  
(t \rightarrow \hbox{hadrons})
\; + \;
( \widetilde{\chi}_2^0 \rightarrow \widetilde{\chi}_1^0 \ell \ell )
b$.
Note though that such processes do have an extra $b$-jet 
(typically with high $p_T$) 
beyond that expected from
$gb \rightarrow (t \rightarrow \hbox{hadrons}) \;
(H^- \rightarrow
\widetilde{\chi}^-_i\widetilde{\chi}^0_j \rightarrow 3 \ell N)$
where $N$ may be any number of colourless neutral stable 
particles. Fortunately, our studies indicate that the extra $b$-jet 
that is present in the $2\rightarrow 3$ charged Higgs boson production
process tends to be rather soft.
So a cut on extra hard jets in the event does tend to remove the
background from stop and sbottom pair production (as well as that from
squark and gluino production in general).

In keeping with the optimal strategy outlined above for the
slepton sector, stops are made heavy to minimise this potential
background.  Thus we deal only with the MSSM backgrounds that must be
present: that from direct ino pair production (since we require
$H^{\pm} \rightarrow$ inos, the inos must be relatively light),
and what coloured-sparticle backgrounds still remain when we have heavy
gluinos and {\sl all} squark inputs pushed up to $1\,\hbox{TeV}$.  
A more in depth study of light stops possibly mimicking our signal 
will be presented in an upcoming analysis \cite{4l}.

\section{mSUGRA Parameter Space}
\begin{figure}[!t]
\begin{center}
\vskip -1.5cm
\epsfig{file=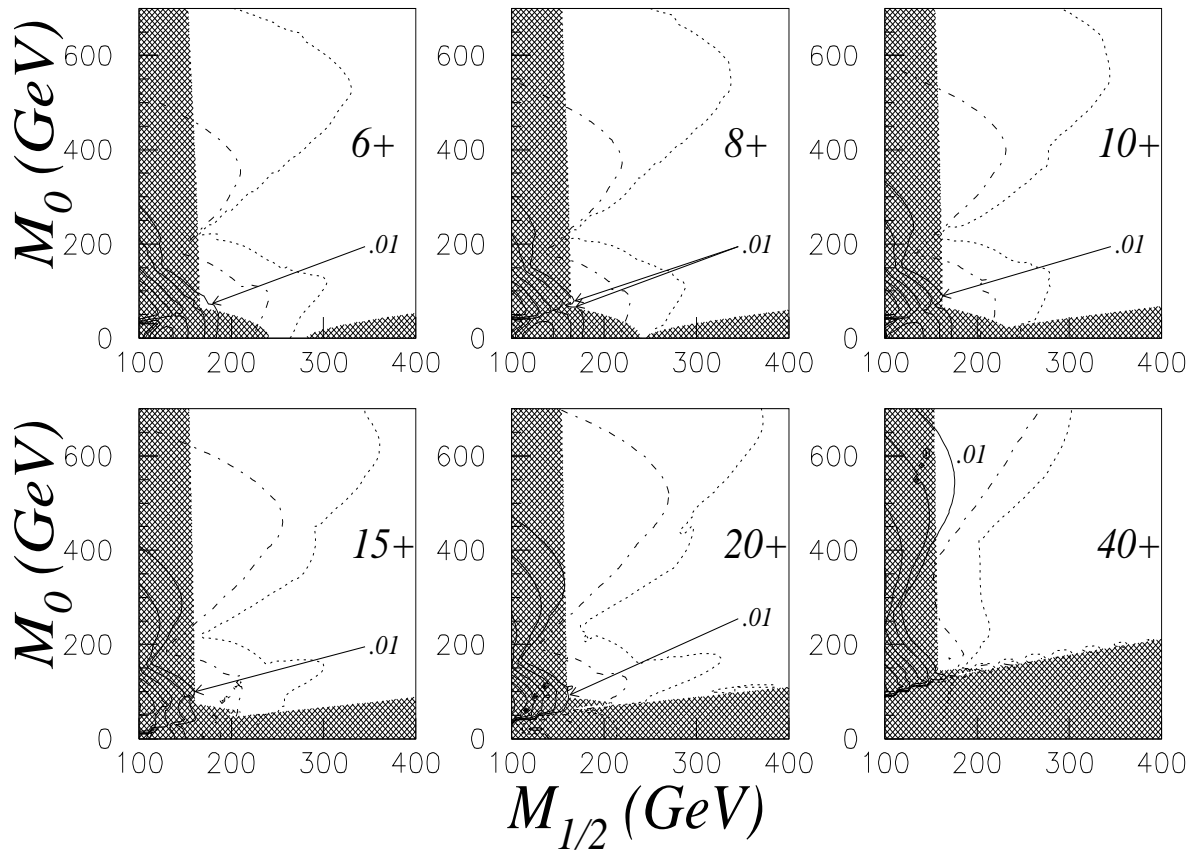,height=140mm, width=170mm}
\end{center}
\vskip -1.25cm
\caption{
$\sigma(pp \rightarrow tH^-X~+~{\rm{c.c.}})$ (in fb) multiplied by
BR$(H^{\pm} \rightarrow 3\ell N)$, where $\ell = e^{\pm}$ or   
${\mu}^{\pm}$
and $N$ represents invisible final state particles,
for a spread of mSUGRA parameter sets in the
$M_{{0}}$ {\it vs.}\  $M_{{\frac{1}{2}}}$ plane;
$A_{{0}} = 0$ in all plots and $\ell$s resulting
from tau decays are included. 
The number in the upper right of each plot is the
$\tan\beta$ value followed by the sign of $\mu$. 
The solid, dot-dashed, and dotted contours are for 
$10^{-2}\, \hbox{fb}$ (also labeled), $10^{-3}\, \hbox{fb}$
and $10^{-4}\, \hbox{fb}$, respectively.  The shaded regions are
excluded by theoretical considerations or LEP2 measurements 
(save that constraints from Higgstrahlung are not applied).
}
\label{fig:mSUGRAHino}
\end{figure}

Before initiating the experimental analysis, we would like 
to document the potential for utilising the 
`$3\ell\, +\, t$ ' signature from $H^{\pm} \rightarrow$ inos 
in the more restrictive mSUGRA parameter space.  
As we will soon see, here prospects are quite bleak.  In mSUGRA, the 
free parameters are generally set as $\tan\beta$,
a universal gaugino mass defined at the Grand Unification Theory (GUT)
scale ($M_{{\frac{1}{2}}}$),
a universal GUT-level scalar mass
($M_{{0}}$),
a universal GUT-level trilinear scalar mass term
($A_{{0}}$),
and the sign of $\mu$.
As already noted, the signal has a strong preference for low values of 
$| \mu |$.  Yet in the mSUGRA scenario, $| \mu |$ is not a free parameter
--- it is closely tied to the masses of the scalar Higgs bosons
via the $M_{{0}}$ input.  Furthermore, the different soft slepton mass
inputs can no longer be set independently:  in particular, 
when evolved down to the EW scale using renormalisation group
equations, the staus' soft inputs tend to be lower than those of sleptons 
from the first two generations rather than higher as was put in
by hand in the more favourable MSSM parameter set choices of the preceding
section.
Fig.\ \ref{fig:mSUGRAHino} shows the values for
$\sigma(pp \rightarrow tH^-X~+~{\rm{c.c.}})$ $\times$
BR$(H^{\pm} \rightarrow 3\ell N)$
obtained for several discrete values of $\tan\beta$ and
$\mu > 0$ (analogous plots for $\mu < 0$ are similar) with $A_0$ set
to zero.
The excluded regions
shown take into account constraints from LEP2 save that coming from
Higgstrahlung\footnote{It should be noted that the small unexcluded
regions shown where the cross section is ${\sim}~0.01\, \hbox{fb}$ 
may be partly or totally excluded by the LEP2 Higgstrahlung constraint.  
However, given the aforementioned uncertainties surrounding this limit,
we conservatively make no attempt to place corresponding exclusion
contours on our plots.
},
but not additional constraints\footnote{See \cite{Batetal} for a
more complete analysis of the present-day constraints on the mSUGRA
parameter space.} from $b \rightarrow s \gamma$, $g_{\mu} - 2$ and other
loop-level effects (nor considerations from cosmology) which
are now harder to dismiss since the behaviour of the model is specified
all the way up to the GUT scale.

In regions of parameter space not LEP2-excluded, maximal rates are found
for very high values of $\tan\beta$.  Here, charged Higgs boson decays 
into three leptons are most favourable when $M_{{0}}$ is low and
$M_{{\frac{1}{2}}}$ is high.  However, the production process (which grows 
with $tan\beta$ once the minimum around $\tan\beta = 6$ is passed) favours 
low $M_{{0}}$ and low $M_{{\frac{1}{2}}}$.  Therefore there is an optimal 
$M_{{0}}$ value for the maximal signal rate (in regions not excluded by
the afore-mentioned LEP2 constraints) which grows from 
${\sim}300\, \hbox{GeV}$ to ${\sim}500\, \hbox{GeV}$ as $\tan\beta$ is 
increased from $20$ to $40$.  Even larger rates for yet higher values of
$\tan\beta$ ($\gsim 60$) are swallowed up by the greatly-expanded
LEP2-excluded region (sparticles or Higgs bosons become unacceptably 
light and/or the lighter stau becomes the LSP).  Inputs must be chosen to
make soft stau masses high (while one would like to --- but in mSUGRA
cannot --- keep the other soft slepton inputs low to obtain good leptonic
decay rates) to avoid a stau LSP and the LEP2 bound on the physical stau
mass.  Thus the window of allowed points for such extremely high values of 
$\tan\beta$ is orthogonal to where substantial signal rates are possible.

Even with $500\, \hbox{fb}^{-1}$ of integrated luminosity, 
the handful of signal events expected in the best allowed cases
would probably be unresolvable from amongst the backgrounds.
Thus, given the very meager chances of extracting a signal even in the
perhaps overly-generous allowed regions of parameter shown here, a
more thorough mSUGRA analysis would probably be irrelevant.

\section{Experimental Analysis}

In the previous sections we outlined the potential for observing the
charged
Higgs bosons through their decays into charginos and
neutralinos, eventually yielding three
leptons plus missing energy, and in the presence of a hadronically 
reconstructed top (anti)quark.
As a next step, we study the feasibility of detecting such a signal in a 
realistic LHC detector environment (CMS). 
We use the MC event generator HERWIG (version 6.3) and simulate 
the $gb \rightarrow tH^{-}~+~{\mathrm {c.c.}} \rightarrow 3\ell +
p_T^{\mathrm{miss}} + t$ signal for the three MSSM settings already
discussed, which we specify more fully here:

\noindent
$\bullet$ Set A: 
$M_{2} = 210\, \hbox{GeV}$, 
$\mu = 135\, \hbox{GeV}$,
$m_{\tilde{\ell}_{\scriptscriptstyle R}} = 110\, \hbox{GeV}$,
$m_{\tilde{g}} = 800\, \hbox{GeV}$, 
$m_{\tilde{q}} = 1\, \hbox{TeV}$.

\noindent
$\bullet$ Set B: 
$M_{2} = 280\, \hbox{GeV}$, 
$\mu = 150\, \hbox{GeV}$,
$m_{\tilde{\ell}_{\scriptscriptstyle R}} = 130\, \hbox{GeV}$,
$m_{\tilde{g}} = 900\, \hbox{GeV}$, 
$m_{\tilde{q}} = 1\, \hbox{TeV}$.

\noindent
$\bullet$ Set C: 
$M_{2} = 300\, \hbox{GeV}$, 
$\mu = -150\, \hbox{GeV}$,
$m_{\tilde{\ell}_{\scriptscriptstyle R}} = 150\, \hbox{GeV}$,
$m_{\tilde{g}} = 1\, \hbox{TeV}$, 
$m_{\tilde{q}} = 1\, \hbox{TeV}$.

\noindent
Recall that in all settings we assume 
$M_{1}
=\frac{5}{3}\tan^2\theta_W M_{2}$. 
Furthermore, for sleptons and squarks we will always take 
soft mass inputs for all generations to be degenerate
(with $m_{\tilde{\ell}_{\scriptscriptstyle L}} =
m_{\tilde{\ell}_{\scriptscriptstyle R}}$).  
The physical sneutrino masses, $m_{\tilde{\nu}}$, can be derived from the
above parameters and are approximately $90$, $115$ and $135\, \hbox{GeV}$
for the respective scenarios (when $\tan\beta \, \gsim \, 5$).
Parameter Set A lies inside the optimal region in the three-dimensional
($M_{2}$, $\mu$, $m_{\tilde{\ell}_{\scriptscriptstyle R}}$) 
space identified in Sect.~2, whereas Set B is a more borderline case
and Set C is a difficult case with a negative $\mu$ parameter.
Set A features light inos and sleptons, allowing several 
supersymmetric $H^{\pm}$ decay modes to have considerable BRs for
relatively moderate values of $m_A$ (and $m_{H^{\pm}}$).      
The ino sectors in Set B and Set C are heavier, thereby limiting the
number of possible sparticle decay modes.  In Set B sleptons are
light, whereas in Set C these sparticles are also heavy.
This last difference markedly alters the kinematics in ways    
to be discussed shortly.
The MSSM sparticle spectrum and decays are obtained from ISASUSY 7.58
\cite{ISAJET} through the ISAWIG interface \cite{ISAWIG}.
ISASUSY contains a one-loop treatment of all Higgs boson masses and
tree-level sfermion masses.
Several three-body decays are included, taking into account 
the full Yukawa contributions, which are important in the large 
$\tan\beta$ regime.
The charged Higgs boson BRs are taken from HDECAY \cite{HDECAY} (again, via
the ISAWIG interface),
which calculates these in accordance with the most recent theoretical
knowledge.
For the SM backgrounds, all leading processes that can produce the
$3\ell + p_T^{\mathrm{miss}} + t$ signature
have been simulated: $t\bar{t}$ ($t\bar{b}W^-$ typically is
1/4 as large), $t\bar{t}Z$, $t\bar{t}\gamma^*$ and $t\bar{t}h$.
Furthermore, all SUSY backgrounds have been considered for the chosen
settings:
ino pair production (including squark+ino production),
squark and/or gluino production
and slepton pair production.
Of these, the first listed class of SUSY contributions has the largest
cross sections in general, because inos are fairly light in comparison to 
the coloured sparticles.
In our scenarios, slepton pair production never results in a three 
lepton final state\footnote{Unless four leptons are produced rather 
than the usual two, and then one lepton is subsequently disregarded due to 
having a $p_T$ value too low to pass our cuts.  
Rates for such events are negligibly small.}, 
so that it will be excluded from further consideration. 
The detector aspects were simulated using CMSJET 4.801 \cite{CMSJET},
which contains fast parametrisations of the CMS detector response and,
for $b$-tagging, a parametrised track reconstruction performance based on
GEANT simulations \cite{CMSIM}.

In Parameter Set A,
the neutralinos $\widetilde\chi^{0}_1$, ${\widetilde\chi}^{0}_2$, 
${\widetilde\chi}^{0}_3$ and ${\widetilde\chi}^{0}_4$
have masses of $78$, $131$, $146$ and $253\, \hbox{GeV}$, respectively.
The masses of the charginos ${\widetilde\chi}^{\pm}_1$ and 
${\widetilde\chi}^{\pm}_2$ are $108$ and $252\, \hbox{GeV}$.
The $H^{\pm}$ are allowed to decay into all kinematically accessible 
ino pairs, ${\widetilde\chi}^{\pm}_i{\widetilde\chi}^{0}_j$,
which in turn can decay into three leptons\footnote{Here $i=1,2$ 
$j=1,2,3,4$; if $j=1$, then the three leptons must all come from cascade 
decays of the chargino.} 
(electrons and/or muons) plus invisible neutral particles
(${\widetilde\chi}^{0}_1$s and/or neutrinos). 
In this scenario,
the primary source of three-lepton events (before any kinematical
cuts are considered) is generally charged Higgs boson decays to
${\widetilde\chi}^{\pm}_1{\widetilde\chi}^{0}_2$.  This is true for
$225\, \hbox{GeV} \, \lsim \, m_A \, \lsim \, 400\, \hbox{GeV}$ 
and\footnote{The upper (lower) $m_A$ value drops by
${\sim} 20\, \hbox{GeV}$ (${\sim} 10\, \hbox{GeV}$)
as $\tan\beta$ goes from $10$ to $30$ ($5$).} $\tan\beta \sim 10$.
Charged Higgs boson decays to
${\widetilde\chi}^{\pm}_2{\widetilde\chi}^{0}_2$
and ${\widetilde\chi}^{\pm}_1{\widetilde\chi}^{0}_4$ are also important
sources of $3\ell$ events in this region of parameter space, and the
contributions from these modes grow to equal or surpass
that from ${\widetilde\chi}^{\pm}_1{\widetilde\chi}^{0}_2$
decays for $m_A \, \gsim \, 400\, \hbox{GeV}$.
${\widetilde\chi}^{0}_2$ decays almost exclusively
via an intermediate state containing an on-shell charged slepton,
while ${\widetilde\chi}^{\pm}_1$ decays through a
intermediate state including an on-shell sneutrino.
Here though BRs for stau decays rise as $\tan\beta$ grows  
(in part due to the fact that, for fixed    
soft slepton mass inputs, the physical mass of the 
${\widetilde\tau}_1^{\pm}$ decreases swiftly as $\tan\beta$ rises to 
higher values):  for $\tan\beta = 5,10,20,30$, the 
${\widetilde\chi}^{0}_2$ BR to staus is about $0.35,0.42,0.61,0.74$,
respectively.  Additionally, about $1/3$ of the ${\widetilde\chi}^{\pm}_1$ 
decays to sneutrinos are to ${\widetilde\nu}_{\tau}$, and
${\widetilde\chi}^{\pm}_1$ decays to
$\widetilde{\tau}_1^{\pm} {\nu}_{\tau}$ become accessible at
$\tan\beta \simeq 11$ with the BR for this decay growing to $0.24$ ($0.06$)
when $\tan\beta$ reaches $30$ ($20$).
These ino to stau and ${\widetilde\nu}_{\tau}$ decays reduce the number of
${\widetilde\chi}^{0}_2 \rightarrow \widetilde{\ell}^{\pm} {\ell}^{\mp}$
and
${\widetilde\chi}^{\pm}_1 \rightarrow \widetilde{\nu}_{\ell} {\ell}^{\pm}$
decays, where ${\ell} = e$ or $\mu$, which lead to virtually all the
signal events that survive the necessary cuts\footnote{If the soft stau
mass inputs are made heavier, rather than degenerate with the other soft
slepton mass inputs as is done in this analysis, the unprofitable
stau channels could be eliminated and the number of events could
as much as double for high values of $\tan\beta$.}.
Leptonic tau decays are allowed in the event generation,
although daughter leptons from these will mostly be rejected during the
analysis stage due to their softness (low $p_T$s).
Crucial mass differences have values\footnote{These numbers depend 
moderately on $\tan\beta$.  Values given here (and later for 
Sets B \& C) cover the range of interest in this work:  
$5 \le \tan\beta \le 30$.  
Physical slepton masses are those of the first two generations.} of 
($m_{{\widetilde\chi}^{0}_2} - m_{\widetilde{\ell}^{\pm}}$,
 $m_{{\widetilde\chi}^{\pm}_1} - m_{\widetilde{\nu}_{\ell}}$,
 $m_{\widetilde{\ell}^{\pm}} - m_{{\widetilde\chi}^{0}_1}$) =
(${\sim}10$-$15\, \hbox{GeV}$,
 ${\sim}10$-$22\, \hbox{GeV}$,
 ${\sim}35$-$45\, \hbox{GeV}$).  
In all these cases there is enough phase space for most of the 
resulting leptons to have sufficiently high transverse momenta.

In order to distinguish between the signal and the backgrounds (both SM
and MSSM), 
we will apply a set of selection criteria that will allow us to obtain a 
favourable signal-to-background ratio using only physically 
well-motivated cuts ({\it i.e.}, with only a very loose dependence upon
the MSSM parameters). 
We will first explain the selection strategy and then illustrate the 
results numerically in a table.

First of all we require the following basic topology:
\begin{itemize}
\item 
\vskip -0.1cm
Events must have exactly three isolated leptons ($\ell = e,\mu$) 
with $p_T > 20, 7, 7\, \hbox{GeV}$, 
all with $|\eta| < 2.4$. The isolation cut demands that there
are no charged particles with $p_T > 1.5\, \hbox{GeV}$ in a cone of radius 
$\Delta R = \sqrt{(\Delta\phi)^2 + (\Delta\eta)^2} = 0.3$ radians 
around each lepton track and that the sum of the
transverse energy deposited in the electromagnetic calorimeter between 
$\Delta R = 0.05$ and $\Delta R = 0.3$ radians be smaller than 
$3\, \hbox{GeV}$.
\end{itemize}

\vskip -0.1cm
The choice of the minimum $p_T$ value for the leptons is driven by both 
trigger and background rejection considerations.
In the case of muons, requiring a hardest lepton above $20\, \hbox{GeV}$ 
is already sufficient for the event to be triggered with 90\% efficiency 
by the single-muon trigger under low luminosity running conditions at the
LHC (for electrons this threshold is somewhat higher) \cite{DAQTDR}. 
Apart from the single leptons triggers, the di- and trilepton thresholds 
will increase the efficiency for triggering on the $3\ell$ signal.
The tight isolation criterion is needed in order to reject leptons coming
from heavy flavour decays, especially in the low $p_T$ region.
As we will discuss later, it is very effective against, 
for instance, the $t\bar{t}$ background, when one or more of the leptons 
originates from a $b$-jet. 


Apart from requiring the three leptons, it is also necessary to reconstruct 
the (hadronically decaying) top quark that is produced in association with 
the $H^{\pm}$ boson.    
This is mainly motivated by the need to strongly suppress 
the $t\bar{t}$ and ino-ino backgrounds.  A reconstructed top 
(antitop) quark is recognised via the following cuts:
\begin{itemize}
\item 
\vskip -0.1cm
Events must have at least three jets, each with 
$p_T > 20\, \hbox{GeV}$ in $|\eta|<4.5$.
\vskip -0.8cm
\item 
Among these, the three jets that are most likely to come
from a top quark decay are selected by minimising $m_{jjj} - m_{t}$, 
where $m_{jjj}$ is the invariant mass of the three-jet system. 
This invariant mass $m_{jjj}$ must be in the range 
$m_t \pm 35\, \hbox{GeV}$.
\vskip -0.8cm
\item 
Two of these three jets are then further
selected by minimising $m_{jj} - M_{W^\pm}$. 
Their invariant mass, $m_{jj}$, must be in the  
range $M_{W^\pm} \pm 15\, \hbox{GeV}$.
\vskip -0.8cm
\item 
The third jet ({\it i.e.}, aside from the two jets 
in the preceding point) must be $b$-tagged.  (This we equate 
with a significance of the transverse impact parameter
$\sigma(ip)=\frac{ip_{xy}}{\Delta ip_{xy}}$ 
which is larger than 2.)
\end{itemize}

\vskip -0.1cm
A strong rejection of $t\bar{t}$ events is obtained after the requirement 
of a hadronically reconstructed top quark 
in addition to the three leptons. Assuming that the three jets
reconstructing $m_t$ are indeed correctly assigned, this requirement 
means that the second top should provide two leptons (one from the $W^\pm$ 
and one from the $b$) while the third lepton should
come from initial/final state radiation ($b,K,\pi, ...$). In this case, two
leptons will be in general soft ($< \, 5\, \hbox{GeV}$) and non-isolated. 
Another scenario in which $t\bar{t}$ production can lead to a $3\ell + t$ 
final state is the one where both top quarks have decayed leptonically,
and two radiated jets accidentally reconstruct the $W^\pm$ mass and
then combine with a $b$-jet from top decay to mimic a
hadronically-decaying top quark.  Here, two leptons can be hard, but the 
third one must still be soft and in general non-isolated.
Therefore, in order to achieve a sufficient suppression of the $t\bar{t}$ 
background, we have chosen to set the lower limit on the $p_T$ of the 
leptons at $7\, \hbox{GeV}$ (although lowering it would increase the 
signal yield) and to apply a tight isolation criterion. 

Whereas the $t\bar{t}$ background is greatly suppressed by the previous 
selection steps, $t\bar{t}Z$, $t\bar{t}\gamma^*$ and $t\bar{t}h$ 
events would still survive the $3\ell+t$ criteria. 
Therefore we require an additional $Z$-veto:
\begin{itemize}
\item 
\vskip -0.1cm
Reject all events with di-lepton pairs
with opposite charges and the same flavour that have an invariant mass 
in the range $M_{Z} \pm 10\, \hbox{GeV}$.
\end{itemize}
\vskip -0.1cm
The $Z$-veto rejects $t\bar{t}Z$ events efficiently. Moreover,
although the $t\bar{t}\gamma^*$ and $t\bar{t}h$ 
backgrounds largely survive this requirement, 
their residual cross sections are now innocuously small.

In addition to eliminating the SM noise, 
cuts to efficiently suppress the SUSY backgrounds that can lead 
to a $3\ell+t$ final state must be considered.
As mentioned before, slepton pair production does not pose a problem 
in our scenarios since it cannot lead to a three-lepton final state.
Ino pair production and squark+ino production can have large
cross sections; however, most events from these processes
do not contain a top quark and will thus be rejected via the 
hadronic top requirement.  Events that are still left after
this cut form the main irreducible SUSY background.
Squark/gluino production is another potentially dangerous source of
noise.   These events, however, typically contain many energetic jets
besides those coming from the top decay (as previously intimated). 
Therefore, they can be rejected using an additional jet veto:
\begin{itemize}
\item 
\vskip -0.1cm
Reject all events containing any jets (other than the three jets
selected for the top reconstruction) with $p_T > \, 70\, \hbox{GeV}$ and
$|\eta|<4.5$.
\end{itemize}

\vskip -0.1cm
For further signal-to-background rejection, we impose the following 
(slightly model dependent) selection criteria (here optimised for 
$m_A = 350\, \hbox{GeV}$ and $\tan\beta = 10$):
\begin{itemize}
\item 
\vskip -0.1cm
For the three isolated leptons already selected,
the $p_T$ of the hardest lepton should be below 
$150\, \hbox{GeV}$ whereas the $p_T$ of the softest lepton 
should be below $40\, \hbox{GeV}$.
\vskip -0.8cm
\item 
The missing transverse energy should be larger than 
$40\, \hbox{GeV}$.
\vskip -0.8cm
\item 
The effective mass, $M_{\mathrm eff}$, constructed from the 
$p_T^{3\ell}$ and $p_T^{\mathrm miss}$ vectors as 
$M_{\mathrm eff} = \sqrt{2 p_T^{3\ell} p_T^{\mathrm miss} 
(1-\cos{\Delta \phi})}$,
is required to be lower than $150\, \hbox{GeV}$ 
(here $\Delta \phi$ is the azimuthal angle between
$p_T^{3\ell}$ and $p_T^{\mathrm miss}$).
\end{itemize}

\vskip -0.1cm
The missing energy requirement has little affect on the signal yield 
since the two ${\widetilde\chi}^0_1$ in the final state
usually supply sufficient $p_T^{\mathrm miss}$; 
on the other hand, this cut does reduce the SM $t\bar{t}V$ 
($V$=$Z$,$\gamma^*$) backgrounds.
As was shown in \cite{PAP1}, the effective mass variable 
does have some dependence on the ino mass spectrum; but it also 
proves to be effective against the above $t\bar{t}V$ processes plus 
squark/gluino and ino pair production backgrounds as well.

After applying these selection criteria, we obtain the
number of signal ($S$) and background ($B$) 
events given in Tab.~\ref{tab:cuts},
assuming Parameter Set A, with $m_A = 350\, \hbox{GeV}$ and 
$\tan\beta = 10$, 
and for an integrated luminosity of $100\, \hbox{fb}^{-1}$.
Results shown therein clearly confirm the points made in the 
preceding description of the cuts. 
\begin{table}[!h]
    \begin{center}
     \begin{tabular}{|l||c|c|c|c|c|c|} \hline
Process   & 3$\ell$ events & $Z$-veto & hadr. top$^{\dagger}$ &
$b$-tag&jet veto$^{\dagger\dagger}$& others$^{\dagger\dagger\dagger}$ \\ 
\hline
$t\bar{t}$             
&    2781     &   2465  &   91   &   15.5  &  11.1   &    5.8        \\
$t\bar{t}Z$          
&     492     &     82  &   19   &    8    &   2.4   &    0.8        \\
$t\bar{t}\gamma^*$     
&      22     &     21  &    7   &    2    &   0.4   &    0.2        \\
$t\bar{t}h$            
&      59     &     52  &   17   &    4    &   1.6   &    0.2        \\
$\widetilde{\chi} \widetilde{\chi}$ 
&   19993     &  18880  &  237   &   31    &   9     &    3          \\
$\tilde{q}$, $\tilde{g}$  
&   12712     &  11269  & 3984   &  861    &   6     &    1          \\    
\hline
$t H^-$ \& $\bar{t}H^+$
&     508     &    485  &  126   &   36    &  29     &   25          \\   
\hline
     \end{tabular}
    \end{center}
\vskip -0.4cm
\caption{Number of signal and background events
assuming Parameter Set A, 
with $m_A = 350\, \hbox{GeV}$ and 
$\tan\beta = 10$, for $100\, \hbox{fb}^{-1}$.
 (Note
that the difference between event rates in the `hadr. top' and the
`$b$-tag' columns is not only due to the
experimental $b$-tagging efficiency but also takes
into account part of the algorithmic efficiency.)
\hskip0.15cm\noindent
$~~~^{\dagger}$Here,  
top reconstruction requires $\ge$ 3 jets with 
$p_T$ $>$ $20\, \hbox{GeV}$, 
$m_{jj} \sim M_{W^\pm}$ and $m_{jjj} \sim m_t$.
\hskip0.15cm\noindent
$~~^{\dagger\dagger}$Here, one vetoes additional jets beyond 3 with 
$p_T$ $>$ $70\, \hbox{GeV}$.
\hskip0.15cm\noindent
$^{\dagger\dagger\dagger}$Here, one imposes 
$p_T(\ell_1)$ $<$ $150\, \hbox{GeV}$, 
$p_T(\ell_3)$ $<$ $40\, \hbox{GeV}$, 
$p_T^{\mathrm{miss}}$ $>$ $40\, \hbox{GeV}$ 
and $M_{\rm eff}$ $<$ $150\, \hbox{GeV}$.
}
\label{tab:cuts}
\end{table}
\begin{figure}[!h]   
\begin{center}
\epsfig{file=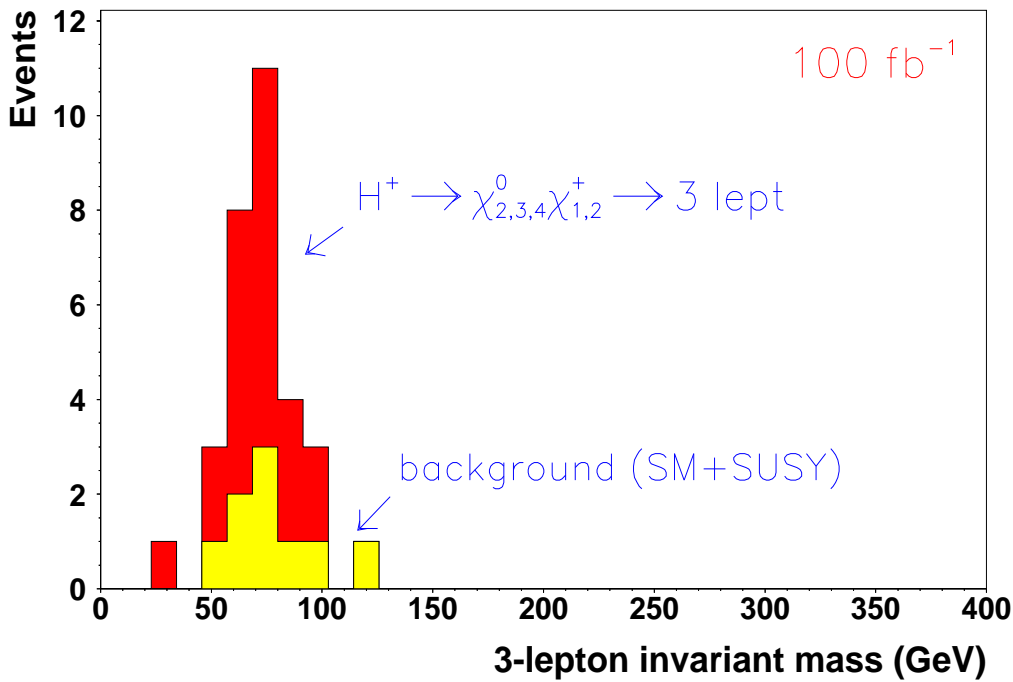,height=100mm, width=140mm}
\end{center}
\vskip -0.3cm
\caption{Three-lepton invariant mass distribution
for Parameter Set A, with $m_A = 350\, \hbox{GeV}$ and
$\tan\beta = 10$. The signal peak is shown on top of the SM + SUSY
background for an integrated luminosity of $100\, \hbox{fb}^{-1}$,
after all cuts described in the text.}
\label{fig:peak}     
\end{figure}  

Fig.\ \ref{fig:peak} shows the three-lepton invariant mass
distribution for our typical signal ($m_A = 350\, \hbox{GeV}$ and
$\tan\beta = 10$) on top of the background (SM + SUSY) for an
integrated luminosity of $100\, \hbox{fb}^{-1}$.
The peak in the three-lepton invariant mass distribution depends both on 
$m_{H^{\pm}}$ {\sl and} on the mass spectrum of the intermediate charginos
and neutralinos.  Therefore, a direct `parameter-independent'
mass reconstruction does not seem feasible at this stage. The determination 
of the charged Higgs mass will require comparisons between the measured
three-lepton invariant mass and MC distributions.


Especially since we cannot claim ability to discern a mass resonance,
it is well to review the strategy of this work.  We attempt to locate
a signal for a charged Higgs boson by comparing the number of events
fitting our criteria {\em at a specific point in the MSSM parameter
space} with and without inclusion of our $gb\to tH^-$ (and c.c.)
subprocess accompanied by $H^{\pm} \rightarrow 3\ell N$.  Events 
resulting from sparticle production processes (sleptons, squarks, 
gluinos, and/or inos) which happen to satisfy our criteria are regarded 
as backgrounds.  It is conceivable that the rate from sparticle production 
processes {\em at a different point in the MSSM parameter space} could 
mimic the excess due to charged Higgs boson production at the
afore-mentioned specific point we are considering.
Thus one could consider instead including the sparticle production
processes in the signal category and mapping out the excess expected
from all non-SM sources throughout the MSSM parameter space.

We have chosen to proceed treating only $H^{\pm}$-events as the
signal events for several reasons:
(1)  Our choice of cuts to select signal events is tailored to pick out
$gb\to tH^-$ (and c.c.), $H^{\pm} \rightarrow 3\ell N$ events;
(2)  We do give only cursory consideration (for the moment) to some rather
limited regions of the parameter space, such as those containing light
sbottoms or stops, which {\em might} yield significant numbers of
non-$H^{\pm}$ signal events (though we do give reasons why we suspect this
will not be the case); and, most importantly, (3) we expect that results
from our channel will be correlated with results in other channels to
resolve most of any ambiguity about the correct location in the MSSM
parameter space.  Sparticle production has other signatures.  For
instance, we cannot restrict our consideration to parameter set choices
where rates for direct ino production are low since we require the charged
Higgs bosons to decay into inos.  But there are other signatures for
direct ino production \cite{directino}, notably trilepton or like-sign
dilepton signals (without an associated $t$ or $\bar{t}$) from
chargino-neutralino production.
In fact, knowledge from these other channels (either from the LHC or
perhaps from runs at the Tevatron) could enable future analyses to sharpen
the cuts employed here (which only incorporate some vague assumed form for
the ino spectrum), and then perhaps reconstruction of the charged Higgs
boson mass will be possible. 
Therefore, our perspective is to treat this work as one piece of a 
body of analyses with which we hope to pin down SUSY's nature.

\begin{figure}[!th]
\begin{center}
\epsfig{file=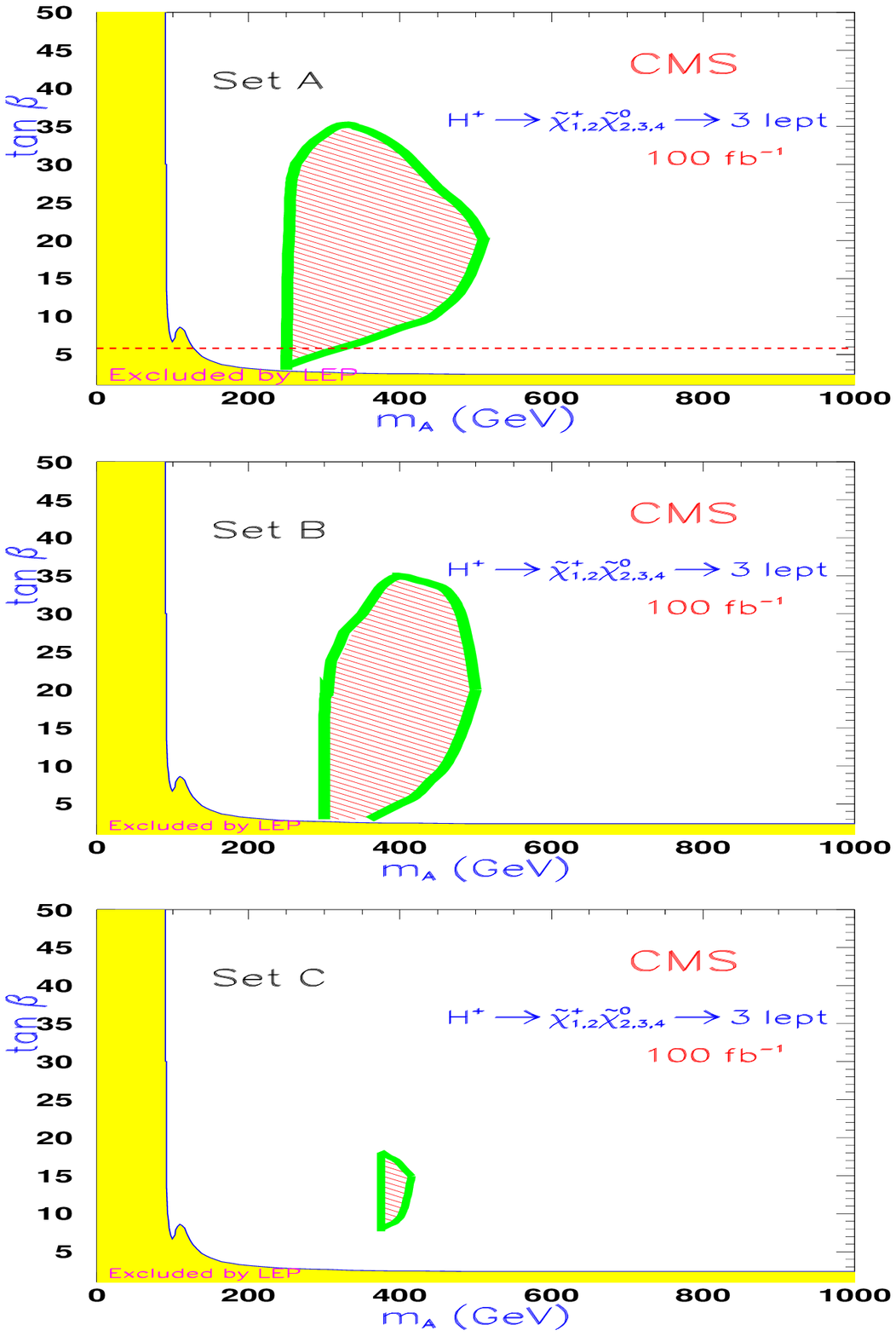, height=185mm, width=130mm}
\end{center}
\vskip 0.1cm
\caption{$5\sigma$-discovery contours
in the $\tan\beta$ {\it vs.}\ $m_A$ plane
for Parameter Sets A, B and C,
assuming an integrated luminosity of $100\, \hbox{fb}^{-1}$.
The shaded region
at the left and bottom of each plot is excluded by LEP2
Higgstrahlung
({\it i.e.},
$e^+e^- \rightarrow h Z$ and $e^+e^- \rightarrow h A$) limits.
The region in the top plot below the
red dotted line is excluded by the LEP2 chargino mass bound.
}
\label{fig:contour}
\end{figure}

Maintaining the MSSM setup of Parameter Set A, we can now perform a scan 
over $m_A$ and $\tan\beta$ in order to determine the discovery potential 
for the `$3\ell + p_T^{\mathrm{miss}} + t$' signature we have been
considering.  We assume an integrated luminosity of $100\, \hbox{fb}^{-1}$
and require the significance of the signal, $S/\sqrt{B}$, to be larger than 
5. The resulting $5\sigma$-discovery potential is shown in the top plot of 
Fig.\ \ref{fig:contour}.
The left edge of the potential discovery region at 
$m_A$ $\approx$ $250\, \hbox{GeV}$ is determined by the kinematic requirement 
that 
$m_{H^{\pm}} > m_{{\widetilde\chi}^{0}_2} + m_{{\widetilde\chi}^{\pm}_1}$. 
The upper edge in $\tan \beta$ originates from decreasing $H^{\pm} \rightarrow
{\widetilde\chi}^{\pm}_i{\widetilde\chi}^0_j$ and ${\widetilde\chi}^0_j 
\rightarrow \ell^+ \ell^- {\widetilde\chi}^0_1$
($i=1,2$ and $j=2,3,4$) BRs.  
This is in part a consequence of the high $\tan \beta$ enhancement of 
$H^{\pm}$ couplings to the third generation (taus, top and bottom quarks),
which grow at the expense of the couplings to the inos
(the intermediates we need to obtain our hard, isolated electrons and
muons) and in part due to the increased BRs for ino decays into staus,
which grow at the expense of decays into lepton($e$ and $\mu$)-yielding 
selectrons and smuons.
The upper edge in $m_A$ and lower edge in $\tan \beta$ are determined by 
the $m_A$ and $\tan\beta$ dependence of the production cross section. 
Conservative LEP exclusion limits \cite{LEPlim}, mainly from Higgstrahlung
({\it i.e.}, $e^+e^- \rightarrow h Z$ and $e^+e^- \rightarrow h A$),
are also drawn in the figure along with a horizontal dotted line below
which $m_{\widetilde{\chi}_1^{\pm}}$ does not respect the LEP2 
bound\footnote{Not shown on the plots in Fig.\ \ref{fig:contour} are
upper $\tan\beta$ bounds of $32.2$, $28.9$, and $44.0$ for 
Parameter Sets A, B and C, respectively, above which the lighter 
stau mass dips below the LSP ($\widetilde{\chi}_1^0$) mass.  
This bound may be evaded by raising the soft stau mass inputs above
those of the first two generations of sleptons.}. 

\centerline{}
\vskip -0.4cm

Parameter Set B produces the following mass spectrum:
the neutralinos ${\widetilde\chi}^{0}_1$, ${\widetilde\chi}^{0}_2$,
${\widetilde\chi}^{0}_3$ and ${\widetilde\chi}^{0}_4$
have masses of $103$, $159$, $165$ and $311\, \hbox{GeV}$, respectively,
while the masses of the charginos ${\widetilde\chi}^{\pm}_1$ and
${\widetilde\chi}^{\pm}_2$ are $131$ and $311\, \hbox{GeV}$.
The $5\sigma$-discovery potential for this setting, again for an
integrated
luminosity of $100\, \hbox{fb}^{-1}$, is presented in the middle plot of
Fig.\ \ref{fig:contour}, after the usual selection procedure.
A noticeable difference with respect to Parameter Set A is that here   
the discovery zone starts at somewhat higher values of $m_A$ due to
the higher $m_{H^{\pm}}$ threshold needed for decays to ino pairs,
since they are heavier than in the previous scenario. 
For $m_A \, \gsim \, 300\, \hbox{GeV}$, 
$H^{\pm} \rightarrow {\widetilde\chi}^{\pm}_1 \widetilde{\chi}_3^0$  
is the dominant source of $3\ell$ events, rather than 
$H^{\pm} \rightarrow {\widetilde\chi}^{\pm}_1 \widetilde{\chi}_2^0$
as in the previous scenario\footnote{For Set B, there is a thin strip of 
parameter space around $m_A \sim 280$-$290\, \hbox{GeV}$ in which 
$H^{\pm} \rightarrow {\widetilde\chi}^{\pm}_1 \widetilde{\chi}_2^0$
is the dominant source of $3\ell$ events.  However, the overall $3\ell$ BR 
drops precipitously in this region as $m_A$ decreases, and so there is 
no potential for discovery.}.
There are two reasons for this change: firstly, with Set B,
$\widetilde{\chi}_3^0$ has a more favourable gaugino/Higgsino mixing 
than does $\widetilde{\chi}_2^0$;
and secondly, for Set B BRs for $\widetilde{\chi}_2^0$ decays into
sneutrino spoiler modes for $\tan\beta = 5,10,20,30$ are about
$85$\%,$84$\%,$68$\%,$39$\%, whereas for Set A these values are  
all roughly $0.5$\%.
$H^{\pm} \rightarrow {\widetilde\chi}^{\pm}_1 \widetilde{\chi}_3^0$
decays remain the dominant source of $3\ell$ events even for high 
charged Higgs boson masses.
The $H^{\pm} \rightarrow {\widetilde\chi}^{\pm}_2{\widetilde\chi}^{0}_2$
and $H^{\pm} \rightarrow {\widetilde\chi}^{\pm}_1{\widetilde\chi}^{0}_4$
decay modes, which led to the majority of the three lepton events for a
${\sim}450\, \hbox{GeV}$ charged Higgs boson with Set A, 
in Set B are no longer dominant for higher $m_{H^{\pm}}$ values, at most 
providing ${\sim}30$\% of the events (before cuts) for 
$m_{H^{\pm}} \sim 650\, \hbox{GeV}$ (for which mass value the production rate 
is already too low for any hope of discovery).
The $\widetilde{\chi}_3^0$ decays predominantly into a charged slepton; the 
BR for $\widetilde{\chi}_3^0$ decays into staus grows from ${\sim}32.5$\% 
to ${\sim}47.5$\% as $\tan\beta$ goes from $5$ to $30$, cutting into the 
desired decays to selectrons and smuons.  Though $\widetilde{\chi}_3^0$
decay modes to sneutrinos are accessible too, the combined BR for such 
spoiler modes remain at the $3$-$4$\% level.
The ${\widetilde\chi}^{\pm}_1$ decays through a sneutrino intermediate 
state and the associated charged lepton\footnote{Chargino decays to 
$\widetilde{\tau}_1^{\pm} {\nu}_{\tau}$
become significant for higher values of $\tan\beta$:
this BR is ${\sim}21$\% (${\sim}6$\%) for $\tan\beta = 30$ ($20$).
Not so useful decays to $\widetilde{\nu}_{\tau} {\tau}^{\pm}$ have BRs
of $34$-$37$\%.}.
Crucial mass differences have values of
($m_{{\widetilde\chi}^{0}_3} - m_{\widetilde{\ell}^{\pm}}$,
 $m_{{\widetilde\chi}^{\pm}_1} - m_{\widetilde{\nu}_{\ell}}$,
 $m_{\widetilde{\ell}^{\pm}} - m_{{\widetilde\chi}^{0}_1}$) =
(${\sim}25$-$30\, \hbox{GeV}$,
 ${\sim}12$-$22\, \hbox{GeV}$,
 ${\sim}30$-$40\, \hbox{GeV}$).
As in Set A, there is enough kinematical phase space for most of the 
resulting leptons to have sufficiently high transverse momenta to pass the 
signal selection criteria.

In Parameter Set C the neutralinos
${\widetilde\chi}^{0}_1$, ${\widetilde\chi}^{0}_2$,
${\widetilde\chi}^{0}_3$ and ${\widetilde\chi}^{0}_4$
have masses of $118$, $162$, $171$ and $324\, \hbox{GeV}$, respectively.
The masses of the charginos
${\widetilde\chi}^{\pm}_1$ and ${\widetilde\chi}^{\pm}_2$ are
$143$ and $324\, \hbox{GeV}$.
Scanning over $m_A$ and $\tan\beta$, after the customary selection 
cuts, now leads to the $5\sigma$-discovery potential seen in the
bottom plot of Fig.\ \ref{fig:contour}.
The reach both in $m_A$ and $\tan\beta$ is strongly reduced
in comparison to the previous scenarios, in part due to the
heavier ino mass spectrum which gives the expected upwards shift
of the left edge in $m_A$.
As with Set B,
$H^{\pm} \rightarrow {\widetilde\chi}^{\pm}_1 \widetilde{\chi}_3^0$
is the dominant source of signal events for
$m_A \, \lsim \, 600\, \hbox{GeV}$ ($m_A \, \lsim \, 520\, \hbox{GeV}$)  
and $\tan\beta \simeq 5$ ($30$).  For
$350\, \hbox{GeV} < m_A < 450\, \hbox{GeV}$, virtually all
($>90$\%) signal events come via this channel.  For higher masses,
$H^{\pm} \rightarrow {\widetilde\chi}^{\pm}_1 \widetilde{\chi}_4^0$
and
$H^{\pm} \rightarrow {\widetilde\chi}^{\pm}_2 \widetilde{\chi}_3^0$
contributions grow to become comparable.
Also, as with Set B, the $\widetilde{\chi}_3^0$ decays predominantly into 
a charged slepton;
again decays into staus --- BR ${\sim}29$\% (${\sim}58$\%)
for $\tan\beta =5$ ($30$) --- cut into the desired decays to selectrons 
and smuons.  The sneutrino spoiler modes also have a combined BR
roughly in the $10$-$20$\% range.
The ${\widetilde\chi}^{\pm}_1$ decays through a sneutrino intermediate 
state: for $\tan\beta < 20$, about $2/3$ of the time into sneutrinos of 
the first two generations and about $1/3$ of the time into a 
${\widetilde\nu}_{\tau}$
and the associated $\tau$-lepton.  For higher values of $\tan\beta$, the
${\widetilde\chi}_1^{\pm} \rightarrow   
\widetilde{\tau}_1^{\pm} {\nu}_{\tau}$
decay mode becomes accesssible and reaches a BR of almost $70$\% by the
time $\tan\beta$ reaches $30$.
Now crucial mass differences have values of
($m_{{\widetilde\chi}^{0}_3} - m_{\widetilde{\ell}^{\pm}}$, 
 $m_{{\widetilde\chi}^{\pm}_1} - m_{\widetilde{\nu}_{\ell}}$,
 $m_{\widetilde{\ell}^{\pm}} - m_{{\widetilde\chi}^{0}_1}$) =
(${\sim}13$-$17\, \hbox{GeV}$,
 ${\sim}4$-$10\, \hbox{GeV}$,
 ${\sim}30$-$45\, \hbox{GeV}$).
Most significantly, there is considerably less phase space available to
leptons produced in chargino to sneutrino decays\footnote{The difference
between $m_{{\widetilde\chi}^{\pm}_1} - m_{\widetilde{\nu}_{\ell}}$
for Set C and the values for Sets A \& B is more striking when
$\tan\beta$ is restricted to be $\ge 10$.  Then for A and B  
$m_{{\widetilde\chi}^{\pm}_1} - m_{\widetilde{\nu}_{\ell}}   
\; {\sim}17$-$22\, \hbox{GeV}$
while for C the value is ${\sim}4$-$7\, \hbox{GeV}$, with this mass
difference growing with increasing $\tan\beta$ for A and B and
shrinking with increasing $\tan\beta$ for C.};
thus, said leptons are typically too soft and usually fail the
$p_T$ cut.
This explains the much smaller discovery reach for Set C compared  
to the one for Set B.

For $\mu < 0$, the same magnitude of $|\mu|$ leads to heavier inos
(in particular, the LSP and lighter chargino).  Thus, for a fixed $|\mu|$,
we expect a smaller signal rate for $\mu < 0$ than for 
$\mu > 0$.  However, the more rapid rise of the chargino mass as $|\mu|$
increases with $\mu < 0$ also means that we can go to smaller $|\mu|$
values on this side before we run afoul of the
LEP2 excluded region\footnote{This is traceable to a term
$\propto 2\mu M_2 \sin 2\beta$ in the formula for the chargino mass,
and hence the asymmetry diminishes as the $\tan\beta$ value increases.}.
Thus, one can shift to lower $|\mu|$ values on the $\mu < 0$ side to
obtain roughly the same rates as found on the $\mu > 0$ side ({\it cf.},
Fig.\ 4 of \cite{PAP1}).

\begin{figure}[!th]
\begin{center}
\vskip -0.4cm
\epsfig{file=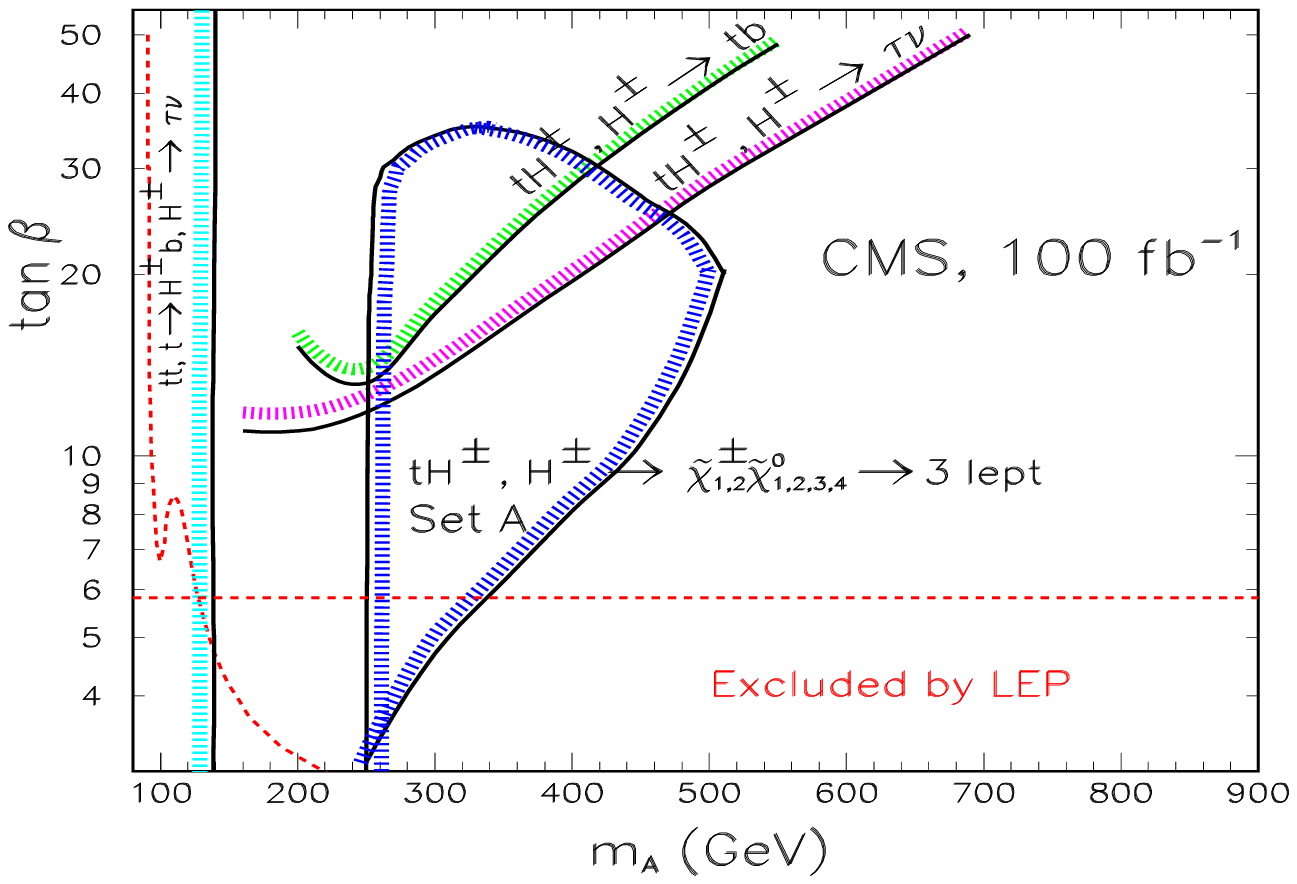,height=120mm, width=140mm}
\end{center}
\vskip -0.5cm
\caption{$5\sigma$-discovery contours 
in the $\tan\beta$ {\it vs.}\ $m_A$ plane
for all charged Higgs channels,
both SM and MSSM,
assuming MSSM inputs as in Parameter Set A and $100\, \hbox{fb}^{-1}$ of
integrated luminosity.
The area below the red dotted line at the left is excluded by LEP2
Higgstrahlung 
({\it i.e.}, $e^+e^- \rightarrow h Z$ and $e^+e^- \rightarrow h A$) 
limits and the region below the horizontal red dotted line is excluded
by the LEP2 chargino mass bound.
}
\label{fig:contour4}
\end{figure}

Some perspective as to the new regions of MSSM parameter 
space that might be probed via the 
`$3\ell + p_T^{\mathrm{miss}} + t$' channel is provided by 
Fig.\ \ref{fig:contour4}, which shows the reach of this
$H^{\pm} \rightarrow$ inos signature 
in the case\footnote{The authors caution that this figure is valid for
a specific set of the MSSM  inputs
$M_{2}$, $\mu$ and
$m_{\tilde{\ell}_{\scriptscriptstyle R}}$, not in general.}
of Parameter Set A
together with those of the $H^-\to\tau^-\bar\nu_\tau$ and 
$H^-\to b\bar t$ channels,
with $\tan\beta$ plotted on a logarithmic scale to better
illustrate the intermediate $\tan\beta$ regime.
The discovery reaches for channels where the $H^{\pm}$ decays to SM particles 
also assume Set A MSSM input parameters and LO normalisation for the
production process; however, said contours do not take into account
possible SUSY backgrounds.
The contour for $H^-\to b\bar t$ also only takes into account the
3$b$-final state analysis \cite{Ritva}. More detailed studies, including
4$b$-final states, are ongoing. However, we do not expect major changes in
the ($m_A,\tan\beta$) reach for this channel.
Similar plots combining the SM and MSSM channels can be drawn for the   
other two MSSM parameter sets.
Comparison of the ``$tb$'' and ``$\tau\nu$'' contours in
Fig.\ \ref{fig:contour4} with the analogous discovery regions in
\cite{CMS}, which used $\mu = -200\, \hbox{GeV}$ and
$M_2 = 200\, \hbox{GeV}$ as inputs, show the former contours to have
shrunk somewhat relative to the latter ones, as expected since  
the combined BR$(H^{\pm} \rightarrow \hbox{inos})$ is larger in relevant
parts of the $(m_A,\tan\beta)$ plane for Set A inputs than for
the inputs of \cite{CMS}.
This shows that the ino decays will reduce the rates for the conventional
$H^{\pm}$ signatures.  In particular, relative to a case where the ino   
decay modes are closed (such as when $|\mu|$, $M_2$, and sfermion masses
are all large) the SM-like discovery regions may be significantly reduced.
This makes the search for the `$3\ell + p_T^{\mathrm{miss}} + t$'
signature from $H^{\pm} \rightarrow \hbox{inos}$ decays all the more
important.

\centerline{}

\section{Conclusions}

In summary, we have proven that SUSY decays of charged Higgs bosons
can profitably be exploited at the LHC in order to detect these
important particles.  We have done an extensive probe of the MSSM
parameter space to see where decays of the type
$H^{\pm} \rightarrow {\widetilde\chi}_i^{\pm}{\widetilde\chi}_j^0$,
($i=1,2$, $j=1,2,3,4$) can yield hadronically quiet three lepton
(electrons and/or muons) final states.  Here all tree-level
decay chains allowable within the MSSM have been taken into account. 
Coupling such decay chains with top-associated charged Higgs
boson production, we selected a signature consisting of 
three hard isolated leptons (electrons and/or muons), 
three hard jets which reconstruct the top quark 
(with one pair thereof also reconstructing a $W^\pm$ boson and the 
other bearing a $b$-tag)
and substantial missing transverse energy.
We then performed quite realistic MC studies utilising the HERWIG event
generator and modelling the CMS detector.  
The hard subprocess used for the signal was $gb\to tH^-$ (and c.c.),
supplemented by initial and final state parton shower and
hadronisation, with overall LO normalisation. (All backgrounds
were generated at the same level of accuracy.)  Recent studies 
\cite{ZhuPlehn} have found that, in contrast to the negative corrections
to said LO production subprocess utilised in the past, NLO corrections
are
in fact positive and may be substantial --- depending upon the choice 
of input scales (see \cite{ZhuPlehn}) $K$-factor enhancements of 
$\gsim \, 1.6$ may be obtained, comparable to or even larger
than the corresponding corrections for the leading backgrounds.
Inclusion of such NLO effects in future
signal and background analyses may well expand the discovery reach of 
this channel.

We found that this `$3\ell + p_T^{\mathrm{miss}} + t$' signature
has the potential to provide coverage over an area of the MSSM
parameter space roughly corresponding to
$250\, \hbox{GeV}$ $<m_{H^{\pm}} < 500\, \hbox{GeV}$ and
$3\,\, \lsim \, \tan\!\beta \,\, \lsim \, 35$.  This region covers a
substantial portion of parameter space where $H^\pm$ decays into ordinary
particles have been shown to be ineffective.
However, to this must be added the {\it caveat} that other MSSM input
parameters must be favourable.  To wit, a small value for $| \mu |$ 
and a small to moderate $M_2$ value are essential for having substantial
$H^{\pm} \rightarrow \widetilde{\chi}_i^{\pm}\widetilde{\chi}_j^0$
BRs (with $M_2 > | \mu |$ to put more weight on ino decays not including
the LSP) and light sleptons are crucial for enhancing the leptonic BRs of
the inos.  Said slepton intermediates may be on- or off- mass shell;
though of course it is optimal if the two-body on-shell ino decay
mode into a slepton and a lepton is open,
as shown by Fig.\ \ref{fig:hto3ell-mslep}.  
Naturally, the actual physical masses of the sleptons (selectrons, smuons
and the associated sneutrinos) should be less than those of an ino pair
into which the charged Higgs boson has a significant BR.
Depending on the ino masses as fixed by the MSSM parameter inputs, this
dictates slepton masses of $\lsim \, 160\, \hbox{GeV}$ or lower in the
discovery regions documented in this work.

Regions in MSSM parameter space satisfying such criteria tend
to be sufficiently close to the LEP2 limits and/or to those derived after 
Run 2 at the Tevatron that such regions should be readily 
accessible to probing by the LHC.
We have made very few assumptions about the underlying SUSY-breaking
dynamics associated with some much higher energy scale, and hence   
defined all relevant MSSM input parameters at the EW scale.
(The mSUGRA model was analysed as a possible GUT benchmark but failed to
shown any potential for the considered decay channel.)
The discovery reach shown in Fig. 9 (for a reasonably favourable   
choice of these parameters) illustrates the possible power of this new
channel.

\section*{Acknowledgments}

MB is grateful to the U.S. National Science Foundation for
support under grant INT-9804704. FM is a Research Assistant of the 
Fund for Scientific Research - Flanders (Belgium).
The authors would like to thank the IUAP/PAI network ``Fundamental 
Interactions'' funded by the Belgian Federal Government.
FM would like to thank Daniel Denegri and Luc Pape for discussions.

\def\pr#1 #2 #3 { {\rm Phys. Rev.}            {#1}, #3 (#2)}
\def\prd#1 #2 #3{ {\rm Phys. Rev. D}          {#1}, #3 (#2)}
\def\prl#1 #2 #3{ {\rm Phys. Rev. Lett.}      {#1}, #3 (#2)}
\def\plb#1 #2 #3{ {\rm Phys. Lett. B}         {#1}, #3 (#2)}
\def\npb#1 #2 #3{ {\rm Nucl. Phys. B}         {#1}, #3 (#2)}
\def\prp#1 #2 #3{ {\rm Phys. Rep.}            {#1}, #3 (#2)}
\def\zpc#1 #2 #3{ {\rm Z. Phys. C}            {#1}, #3 (#2)}
\def\epjc#1 #2 #3{ {\rm Eur. Phys. J. C}      {#1}, #3 (#2)}
\def\mpl#1 #2 #3{ {\rm Mod. Phys. Lett. A}    {#1}, #3 (#2)}
\def\ijmp#1 #2 #3{{\rm Int. J. Mod. Phys. A}  {#1}, #3 (#2)}
\def\ptp#1 #2 #3{ {\rm Prog. Theor. Phys.}    {#1}, #3 (#2)}
\def\jhep#1 #2 #3{ {\rm J. High Energy Phys.} {#1}, #3 (#2)}
\def\jphg#1 #2 #3{ {\rm J. Phys. G}           {#1}, #3 (#2)}

\end{document}